%% file: mnras_template.tex
\documentclass[fleqn,usenatbib]{mnras} 

\usepackage{newtxtext,newtxmath}

\usepackage[T1]{fontenc}
\usepackage{lipsum}
\usepackage{comment}

\DeclareRobustCommand{\VAN}[3]{#2}
\let\VANthebibliography\thebibliography
\def\thebibliography{\DeclareRobustCommand{\VAN}[3]{##3}\VANthebibliography}

\usepackage{graphicx}	
\usepackage[dvipsnames]{xcolor}
\usepackage{mathrsfs}

\title[AGN and precipitation of the ICM]{The interplay between AGN feedback and precipitation of the intracluster medium in simulations of galaxy groups and clusters}

\author[F.S.J. Nobels et al.]{
Folkert S.J. Nobels,$^{1}$\thanks{E-mail: nobels@strw.leidenuniv.nl}
Joop Schaye$^{1}$,
Matthieu Schaller$^{1,2}$,
Yannick M. Bahé$^{1}$ 
and Evgenii Chaikin$^{1}$
\\
$^{1}$Leiden Observatory, Leiden University, PO Box 9513, NL-2300 RA Leiden, The Netherlands\\
$^{2}$Lorentz Institute for Theoretical Physics, Leiden University, PO Box 9506, NL-2300 RA Leiden, The Netherlands
}

\date{Accepted XXX. Received YYY; in original form ZZZ}

\pubyear{2022}

\begin{document}
\label{firstpage}
\pagerange{\pageref{firstpage}--\pageref{lastpage}}
\maketitle

\begin{abstract}
\input{abstract.tex}
\end{abstract}




\section{Introduction}
\label{sec:introduction}

\input{intro.tex}

\section{Initial conditions} 
\label{sec:IC}
\input{IC.tex}

\input{simulations.tex}

\section{Results}
\label{sec:results}

\input{initial_phase.tex}

\input{regulation.tex}

\section{Discussion}
We continue with a discussion of our results in the context of other similar studies. Specifically, we look at the connection to the precipitation framework (\S \ref{subsec:precip}), numerical resolution (\S \ref{subsec:resolution}) and comparisons with other simulations (\S \ref{subsec:previous-sims}). 

\label{sec:discussion}

\input{precipitation.tex}

\input{resolution.tex}

\input{init_profile}
\input{comparison}

\input{conclusion}

\section*{Acknowledgements}
The research in this paper made use of the SWIFT open-source simulation code (\url{http://www.swiftsim.com}, \citealt{schaller2018}) version 0.9.0. This work used \texttt{swiftsimio} \citep{borrow2020b} for reading and visualising the data, for SPH projections the subsampled projection backends were used \citep{borrow2021}. This work used the DiRAC@Durham facility managed by the Institute for Computational Cosmology on behalf of the STFC DiRAC HPC Facility (www.dirac.ac.uk). The equipment was funded by BEIS capital funding via STFC capital grants ST/K00042X/1, ST/P002293/1, ST/R002371/1 and ST/S002502/1, Durham University and STFC operations grant ST/R000832/1. DiRAC is part of the National e-Infrastructure.
This work was suppored by the Netherlands Organization for Scientific Research
(NWO) through Veni grant number 639.041.751 and Vici grant 639.043.409 and by the European Union’s Horizon 2020 research and innovation programme under the Marie Sklodowska-Curie grant agreement No 860744 (BiD4BESt).

\section*{Data Availability}

The initial conditions of the simulations presented in this paper will be made publicly available in \texttt{SWIFT}. \texttt{SWIFT} is publicly available and can be found at \url{www.swiftsim.com}. The simulations can be reproduced by running \texttt{SWIFT} with these initial conditions. The other data underlying this article will be shared on reasonable request to the corresponding author.



\bibliographystyle{mnras}
\bibliography{mnras_template} 




\appendix

\input{seed.tex}


\bsp	
\label{lastpage}
\end{document}

%% file: abstract.tex
\noindent Using high-resolution hydrodynamical simulations of galaxy clusters, we study the interaction between the brightest cluster galaxy, its supermassive black hole (BH) and the intracluster medium (ICM). We create initial conditions for which the ICM is in hydrostatic equilibrium within the gravitational potential from the galaxy and an NFW dark matter halo. Two free parameters associated with the thermodynamic profiles determine the cluster gas fraction and the central temperature, where the latter can be used to create cool-core or non-cool-core systems. Our simulations include radiative cooling, star formation, BH accretion, and stellar and active galactic nucleus (AGN) feedback. Even though the energy of AGN feedback is injected thermally and isotropically, it leads to anisotropic outflows and buoyantly rising bubbles. We find that the BH accretion rate (BHAR) is highly variable and only correlates strongly with the star formation rate (SFR) and the ICM when it is averaged over more than $1~\rm Myr$. We generally find good agreement with the theoretical precipitation framework. In $10^{13}~\rm M_\odot$ haloes, AGN feedback quenches the central galaxy and converts cool-core systems into non-cool-core systems. In contrast, higher-mass, cool-core clusters evolve cyclically. Episodes of high BHAR raise the entropy of the ICM out to the radius where the ratio of the cooling time and the local dynamical time $t_{\rm cool}/t_{\rm dyn} > 10$, thus suppressing condensation and, after a delay, the BHAR. The corresponding reduction in AGN feedback allows the ICM to cool and become unstable to precipitation, thus initiating a new episode of high SFR and BHAR.

%% file: intro.tex
Galaxy groups and clusters are the most massive collapsed structures in the Universe and are unique probes of the large-scale structure because the hot gas in the intracluster medium (ICM) can be observed in X-rays and through the Sunyaev-Zel’dovich effect (SZE) \citep[][for a review]{kravtsov2012}. Furthermore, galaxy clusters are extreme environments where star formation in the central galaxy is suppressed by feedback from the supermassive black hole (BH). Therefore, galaxy clusters can be used to study the interplay between the BH and the ICM and to understand its role in galaxy formation more broadly.

X-ray observations indicate that there are two types of galaxy groups and clusters: those with and without a `cool core' of gas with low temperature, low entropy, high density, and short cooling times \citep[e.g.][]{hudson2010}. These two classes are typically referred to as `cool-core' (CC) and `non-cool-core' (NCC) clusters, respectively. The precise ratio of CC to NCC clusters remains a matter of debate, complicated by a bias of X-ray selected samples towards the (X-ray brighter) CC type \citep[e.g.][]{lin2015,rossetti2017}. Samples of SZE-selected massive galaxy clusters indicated that --- $\approx 30-40$ per cent of clusters at redshift $z \leq0.35$ are CC \citep{andrade-santos2017}. For their (statistical) progenitors at $z=0.3-1.3$, \citet{ruppin2021} derived a CC fraction of $\approx 50$ per cent, indicating that the fraction of CC clusters remains almost constant with increasing redshift. 

The observed thermodynamic profiles of CC clusters would suggest the presence of substantial cooling flows, in which large amounts of gas cool (with rates from $10$s to $1000$s $\rm M_\odot ~\rm yr^{-1}$). However, the observed rate at which gas turns into stars is generally much lower, otherwise the brightest cluster galaxies (BCGs) of galaxy groups and clusters would e.g. be far bluer than observed, an inconsistency that has been termed the `cooling flow problem' \citep{fabian1994}. In addition X-ray spectra (see \citealt{gitti2012} for a review) suggest that the amount of gas that is actually able to cool radiatively to low temperatures ($T\la 10^4~\rm K$) is a factor 2-10 lower than expected from the thermodynamic profiles alone. This indicates that the cooling flows in galaxy groups and clusters, where they exist, must typically be truncated close to the BCG\footnote{Note that there are some BCGs with SFRs of 10s to 100s of $\rm M_\odot ~\rm yr^{-1}$, especially at high redshifts.}. It is generally thought that this is accomplished by an active galactic nucleus (AGN) powered by an accreting BH in the centre of the BCG, which heats the gas and thus offsets the radiative cooling \citep[e.g.][]{mcnamara2007, fabian2012,gitti2012,werner2019,eckert2021}. This causes the gas cooling time ($t_{\rm cool}$) to increase and the cooling flow to stop, with the gas regulating to $t_{\rm cool}/t_{\rm dyn} \approx 10$, where $t_{\rm dyn}$ is the dynamical time (i.e. the time in which the gas would fall to the halo centre in the absence of pressure support; \citealt{voit2015,babyk2018}).

Direct observational evidence for AGN feedback is abundant, especially in X-rays: for instance shocks, ripples and cavities have been detected \citep[see][for a review]{eckert2021}, at scales from groups up to massive galaxy clusters. The X-ray cavities, lower surface brightness features caused by higher temperature and lower density, are thought to be formed due to the injection of energy by the AGN \citep[e.g.][]{eckert2021}. This creates high-entropy bubbles that rise buoyantly and expand, giving rise to larger sizes at greater distances from the centre of the BCG \citep[e.g.][]{dong2010}. Deep observations of X-ray cavities indicate that the power expended to inflate them ($P_{\rm cav}$) can balance the cooling-induced gas luminosity ($L_{\rm cool}$), with a very tight $P_{\rm cav}- L_{\rm cool}$ relation ranging from isolated elliptical galaxies to galaxy clusters \citep[e.g.][]{birzan2008,cavagnolo2010,o-sullivan2011,hlavacek-larrondo2012,eckert2021}. 

However, some cold gas is observed close to the centre of most CC galaxy groups and clusters, implying that cooling flows are not stopped completely \citep[e.g.][]{eckert2021}. Observed cold gas masses range from none \citep[e.g.][]{salome2003,pulido2018} to relatively small amounts ($\approx 3\times 10^8~\rm M_\odot$; \citealp{salome2003,pulido2018}) and to substantial reservoirs ($1.8\times 10^{11}~\rm M_\odot$; \citealp{edge2001}). Resolved cold gas observations often reveal a filamentary structure \citep[e.g.][]{salome2006, salome2011, david2014, temi2018, olivares2019, russell2019}, as well as interactions with satellite galaxies that transport cold gas to the centre of the BCG \citep[e.g.][]{dasyra2012}. Observations indicate that most of the cold gas likely cooled from the hot ICM and is similar for isolated elliptical galaxies and galaxy clusters \citep[see][for a review]{werner2019}.

For lower-mass haloes ($M_{200} \la 3\times 10^{13}~\rm M_\odot$, where $M_{\Delta}$ is the mass within the radius $R_{\Delta}$ where the average internal density $\langle \rho_{} \rangle = \Delta ~\rho_{\rm crit}$ and $\rho_{\rm crit}$ is the critical density of the Universe), detailed X-ray observations are challenging \citep[e.g.][]{eckert2021}. However, even without direct observational constraints on the presence of AGN at these mass scales, cosmological hydrodynamical simulations need to include AGN feedback in order to predict realistic star formation rates (SFR), stellar masses at $M_\star \ga 10^{11}~\rm M_\odot$ and gas fractions inside $R_{500}$ in haloes with mass $M_{200} > 10^{12}~\rm M_\odot$ \citep[e.g.][]{booth2009, mccarthy2010,crain2015, dave2019}. The same holds for non-cosmological simulations of clusters \citep[e.g.][]{li2014a}.

Simulations of group/cluster haloes have great potential to provide understanding of the interplay between the BH and ICM through AGN feedback. Many theoretical studies have investigated different aspects of this interplay in the past. Simulations without any AGN feedback generally develop stable cooling flows that build up unphysically massive gas disks \citep[e.g.][]{li2012}. \citet{gaspari2011a} and \citet{meece2017} investigated different accretion models like Bondi-Hoyle or cold gas triggered accretion. Subsequently, \citet{gaspari2014} and \citet{meece2017} investigated different implementations of energy injection for AGN feedback. These studies indicate that when AGN feedback is self-regulated, run-away cooling is prevented, and cold masses and radial profiles agree with observations. The exact mechanism of BH accretion is then not important.

Simulations that focused more closely on the interaction between the ICM and the BH found that when thermal instabilities grow, such that $t_{\rm cool}/t_{\rm dyn} \lesssim 10$ locally, gas starts to condense, loses its pressure support, and falls towards the centre of the halo \citep[e.g.][]{mccourt2012,sharma2012,gaspari2012,gaspari2013}; this process is often referred to as `precipitation' \citep{voit2015b}. The perturbations can be seeded by AGN feedback due to the lifting of colder gas to regions of larger $t_{\rm dyn}$ \citep[e.g.][]{gaspari2013, li2014b, li2014a}, or by the accretion of cold gas with $t_{\rm cool}$ below equilibrium. However, when $t_{\rm cool}/t_{\rm dyn} > 10$, any radial oscillations are damped because the gas is cooling slowly and remains in a quasi-steady state. Consequently, the ICM is expected to self-regulate towards $t_{\rm cool}/t_{\rm dyn} \approx 10$ on timescales of $\sim 1~\rm Gyr$. Based on this general framework, \citet{voit2017} concluded that gas starts to precipitate when $\alpha_{\rm K}\equiv \frac{{\rm d} \ln K }{{ \rm d} \ln r} < \left( \frac{t_{\rm dyn}}{t_{\rm cool}} \right)^2$, where $K$ is the entropy.

At the same time, cosmological simulations have given clues to the overall impact of AGN feedback on the baryon content of massive haloes \cite[e.g.][]{lebrun2014,barnes2017,barnes2018,weinberger2017, pillepich2018,henden2018,zinger2018,dave2019,davies2019}. Detailed insight into the ICM-AGN connection, however, necessitates higher resolution than is typically affordable in this set-up, and even when cosmological simulations have reached sub-kpc resolution \citep[e.g.][]{pillepich2019, tremmel2019}, these are limited to small samples run with one particular model only. Simulations of idealised galaxy groups/clusters allow detailed experiments with careful control of individual model parameters and initial conditions. 

In this paper, we study the interaction between the ICM and the central BH in galaxy groups and clusters with a suite of high-resolution ($m_{\rm gas}=10^5~\rm M_\odot$) idealised simulations of haloes in the mass range $M_{200}=10^{13}-10^{14}~\rm M_\odot$. All our simulations include an AGN model based on a modified Bondi-Hoyle accretion prescription, with feedback energy injected in thermal form. For a self-consistent treatment of gas cooling, star formation, stellar and AGN feedback we use a subgrid model similar to that employed in the EAGLE project \citep{schaye2015,crain2015}. As we will show, this set-up yields realistic NCC or CC clusters (depending on the initial central temperature), whose cooling times are regulated around $1~\rm Gyr$ without unrealistic cooling flows.

\raggedbottom
The remainder of this paper is structured as follows. In Section \ref{sec:IC} we explain how we set up our idealised galaxy cluster simulations. This is followed by an overview of our subgrid-physics model in Section \ref{sec:subgrid} and a description of our simulations and the results obtained from them in Section \ref{sec:results}. Our results are compared with other theoretical models and observations in Section \ref{sec:discussion}. In Section \ref{sec:conclusion} we summarise our main conclusions.

%% file: IC.tex
\defcitealias{navarro1997}{NFW}

Our simulation set-up consists of a \citet{navarro1997} dark matter halo (hereafter \citetalias{navarro1997}), a central elliptical galaxy with a \citet{hernquist1990} stellar profile, a single BH at the centre, and a gaseous halo in hydrostatic equilibrium as described below, that is simulated using smoothed-particle hydrodynamics (SPH). We limit our study to halo masses of $M_{200}=10^{13}~\rm M_\odot$, $10^{13.5}~\rm M_\odot$ and $10^{14}~\rm M_\odot$, spanning the range from groups to low-mass clusters, which offers the best compromise between computational feasibility and available observations \citep[see][for a review]{oppenheimer2021}. We achieve a very high resolution ($10^5~\rm M_\odot$ in mass for the gas and stars, $300~\rm pc$ for the gravitational softening) in the central 100 kpc of our simulation. This is comparable to the studies of \citet{qiu2019} or \citet{meece2017}, and at least an order of magnitude higher than in \citet{li2015}, \citet{gaspari2014}, and in state-of-the-art cosmological galaxy formation simulations like EAGLE \citep{schaye2015} and TNG100 \citep{weinberger2017,pillepich2018}. Recently the TNG50 \citep{nelson2019, pillepich2019} and RomulusC \citep{tremmel2019} simulations have modelled comparable massive haloes at similar resolution but as discussed above, our idealised set-up allows a cleaner and more straightforward interpretation of the simulation results.

\subsection{Dark matter halo}
For the collisionless dark matter and stars we follow closely the approach of \citet{hernquist1993} and \citet{springel1999}. To limit computational expense, the dark matter halo is represented by a static external potential corresponding to an \citetalias{navarro1997} profile,
\begin{equation}
    \Phi_{\rm DM}(r) = - \frac{4\pi G \rho_0 R_{\rm 200}^3}{c^3 r} \ln \left( 1 + \frac{c r}{R_{\rm 200}} \right).
\end{equation}
where $G$ is the gravitational constant and $r$ the cluster-centric radius. The characteristic density $\rho_0$ depends on the radius $R_{200}$, its enclosed mass $M_{200}$, and its concentration $c$ as $\rho_0 = M_{200} \,c^3/4 \pi R_{200}^3 \left[ \ln(1+c) - \frac{c}{1+c} \right]$. For each halo mass the concentration-mass relation of \citet{correa2015} is used to calculate $c$ with the \citet{planck2016} cosmology.

\subsection{Stars}
To generate the stars, we use a modified version of \texttt{MakeNewDisk} \citep[][]{springel2005}. The stellar component of our elliptical galaxy is modelled as a radially symmetric sphere corresponding to a \citet{hernquist1990} density profile,
\begin{align}
    \rho_{\star} &= \frac{M_\star}{2 \pi} \frac{r_\star}{r(r_\star + r)^3}.
\end{align}
This distribution has two free parameters, the total stellar mass $M_\star$ and scale length $r_\star$, where the latter is set by hand to a specific value depending on the halo mass (see Table \ref{tbl:extra_params}). We do not include any satellite galaxies in our simulations.

While it is straightforward to sample the positions of star particles\footnote{This can be done analytically and requires no artificial truncation, 90 (95) per cent of the mass is within $18 r_\star$ ($38 r_\star$).}, assigning particle velocities such that the system is in dynamical equilibrium requires more care. Following the approach of \citet{hernquist1993} and \citet{springel1999}, we assume that at a specific position the velocity distribution corresponds to the solution of the collisionless Boltzmann equation, and can be approximated sufficiently accurately by a multivariate Gaussian. We can then obtain the moments by using the generalized Jeans equations derived by \citet{magorrian1994} for an axisymmetric system (since we include angular momentum in the gas, see below). These equations constrain the velocity structure such that most first and second moments of the multivariate Gaussian are equal to zero. The only non-zero components are $\langle v_{\rm z}^2 \rangle$, $\langle v_{\rm R}^2 \rangle$, $\langle v_\phi^2 \rangle$ and $\langle v_\phi \rangle$:
\begin{align}
 \langle v_{\rm z}^2 \rangle = \langle v_{\rm R}^2 \rangle = \frac{1}{\rho} \int\limits_z^\infty \rho_\star (z', R) \frac{\partial \Phi_{\rm{total}}}{\partial z'} \mathrm{d} z', \label{eq:vz2}
\end{align}
where $\rho_\star$ is the stellar density, $R$ the radius perpendicular to the $z$-axis and $\Phi_{\rm{total}}$ the total gravitational potential of the system;
\begin{align}
 \langle v_{\rm \phi}^2 \rangle = \langle v_{\rm R}^2 \rangle + \frac{R}{\rho} \frac{\partial (\rho \langle v_{\rm R}^2 \rangle )}{\partial R} + v_{\rm c}^2, \label{eq:vphi}
\end{align}
where $v_{\rm c}^2 = R \partial \Phi_{\rm total}/\partial R$ and $\langle v_\phi \rangle$ is the mean angular streaming component, which can be freely chosen in principle but is set to zero in our models such that the galaxy has no net rotation. Each stellar particle is assigned an initial metallicity of $Z_\star = 1/3 ~Z_\odot$ (where $Z_\odot=0.0134$, \citealt{asplund2009}), a stellar mass of $m_\star=10^5~\rm M_\odot$, and a stellar age of $9 ~\rm Gyr$.

\subsection{Intracluster gas}
After constructing the stellar and dark matter components, the next step is determining the structure of the gas component. We assume that the gas is initially in hydrostatic equilibrium,
\begin{align}
    \frac{{\rm d} P}{{\rm d} r} &= - \frac{G M_{\rm encl}(r) \rho(r)}{r^2}. \label{eq:he}
\end{align}
For simplicity, we neglect the contribution of the gas to the gravitational potential given its subdominant density \citep[e.g.][]{schaller2015}, i.e. $M_{\rm encl}(r)=M_{\rm DM, \rm encl} (r) + M_{\star, \rm encl} (r)$. It is convenient to rewrite equation (\ref{eq:he}) in terms of the circular velocity $v_{\rm c}$ and sound speed $c_{\rm s}$ \citep[e.g.][]{stern2019}:
\begin{equation}
    \frac{{\rm d} \ln P}{\mathrm{d} \ln r} = - \gamma \frac{v_{\rm c}^2}{c_{\rm s}^2}, \label{eq:hydrostatic}
\end{equation}
where $\gamma=5/3$ is the adiabatic index of the gas. Under the assumption of constant $v_{\rm c}/c_{\rm s}$ this corresponds to
\begin{align}
    P = P_{0} \left(\frac{r}{r_{0}}\right)^{-\gamma}, \label{eq:pressure-large-radii}
\end{align}
where $P_{0}$ is a free normalisation pressure at radius $r_{0}$. Gravitationally bound solutions have $v_{\rm c} \sim c_{\rm s}$ \citep[e.g.][]{stern2019}. Assuming that $v_{\rm c}=c_{\rm s}$, the definitions of $v_{\rm c}$ and $c_{\rm s}$ yield a temperature profile 
\begin{align}
   T_{\rm circ} &= \frac{\mu m_{\rm p}}{k_{\rm B} \gamma}  \frac{G M_{\rm encl} (r)}{r}, \label{eq:temp-rotation}
\end{align}
where $m_{\rm p}$ is the proton mass, $k_{\rm B}$ is the Boltzmann constant and $\mu=0.6$ is the mean particle mass for fully ionized gas with a metallicity of $Z=1/3 Z_\odot$. However, the assumption that $v_{\rm c} =c_{\rm s}$ breaks down in the centre of most galaxies because the gas is heated by feedback. To make our initial gas component more realistic, we therefore impose a temperature floor near the halo centre with the gas temperature given by
\begin{align}
    T_{\rm total} &= T_{\rm circ} + \frac{T_{\rm 0}}{1 + \exp\left(\frac{r-2 r_{\rm 0}}{r_{\rm 0}}\right)},
\end{align}
with two free parameters $T_0$ and $r_0$ (see below). We choose this functional form as it asymptotes smoothly to $T_{\rm circ}$ at large radii and to a constant value of $T_{\rm min} \approx 0.9 T_{\rm 0}$ at $r \ll r_0$. With this modified temperature profile, it is no longer possible to solve equation (\ref{eq:hydrostatic}) analytically but we solve it numerically, and as expected the solution approaches equation (\ref{eq:pressure-large-radii}) at large radii. Our gaseous halo extends out to $3 R_{200}$.

The gas distribution in our initial conditions depends on a free parameter $T_0$ which specifies the temperature plateau near the halo centre and can be used to create a CC or NCC system. We calibrate  its value by comparing our initial gas profile against the BAHAMAS simulation \citep{mccarthy2017}, a self-consistent cosmological hydrodynamical simulation at much lower resolution ($m_{\rm gas} \approx 10^9~\rm M_\odot$), as described in \S \ref{sec:BAHAMAS}. As shown by \citet{mccarthy2017}, BAHAMAS matches the radial density and pressure profiles of observed galaxy clusters very well and therefore provides an ideal reference for our idealised simulations. To set the normalisation of our gas density profile, we find the median gas fraction $f_{500} = M_{\rm gas}/M_{500}$ at $z=0$ and adjust $P_{0}$ to match $f_{500}$. 

To set the initial (bulk) velocity of the ICM, we make the simplifying assumption that the angular momentum of the gas traces that of the dark matter halo, with the rotation vector aligned with the $z$-axis. We use the DM angular momentum distribution and radial dependence within $R_{200}$ from \citet{bullock2001}, but with a slightly different spin parameter of the gas ($\lambda = 0.05$) based on \citet{oppenheimer2018}. As we did for stars, we assign each gas particle an initial metallicity of $1/3~{\rm Z_\odot}$ \citep[e.g.][]{werner2013, mcdonald2016}.

\subsubsection{Radially degrading mass resolution} \label{subsubsec:degrading}
Most of the gas mass is at $r \gg 100~\rm kpc$ and is therefore not directly involved in the interplay between gas cooling and feedback. However, our model becomes increasingly unrealistic at these large radii due to the absence of cosmological accretion and satellites. To minimize the computational expense, we therefore only sample the gas at our target resolution ($m_{\rm gas}=m_\star=10^5~\rm M_\odot$ for our fiducial resolution) within $r_{\rm highres} =100~\rm kpc$. At larger radii, the gas particle mass increases as $m_\star ( r/r_{\rm highres} )^2$, reaching $9 \times 10^5~\rm M_\odot$ at $r=300~\rm kpc$ and $4.2 \times 10^6~\rm M_\odot$ at $r=650~\rm kpc$ (corresponding to $R_{500}$ for our most massive halo). This approach is similar to what has been done on smaller scales \citep[e.g.][]{vdvoort2019}. We have verified that using this gradually degrading resolution does not influence the evolution within $r_{\rm highres}$, by running a set of non-radiative simulations with different resolution profiles. Low-resolution particles do not enter the inner $\approx90~\rm kpc$ for the entire duration of our simulations. 

\subsection{The central supermassive black hole}
A single BH is placed in the centre of the galaxy, with a mass given by the BH mass-stellar mass relation from \citet{mcconnell2013}. To avoid an unrealistically high accretion and feedback at the very start of the simulation, we reduce its ambient gas density by removing any gas particles within $1~\rm kpc$\footnote{This corresponds to less than $10^{-4}$ of the gas within $R_{500}$.} of the BH but still keep the total gas mass within $R_{500}$ fixed by placing the removed gas particles randomly between $1~\rm kpc$ and $R_{500}$.

\begin{figure*}
	\includegraphics[width=.97\linewidth]{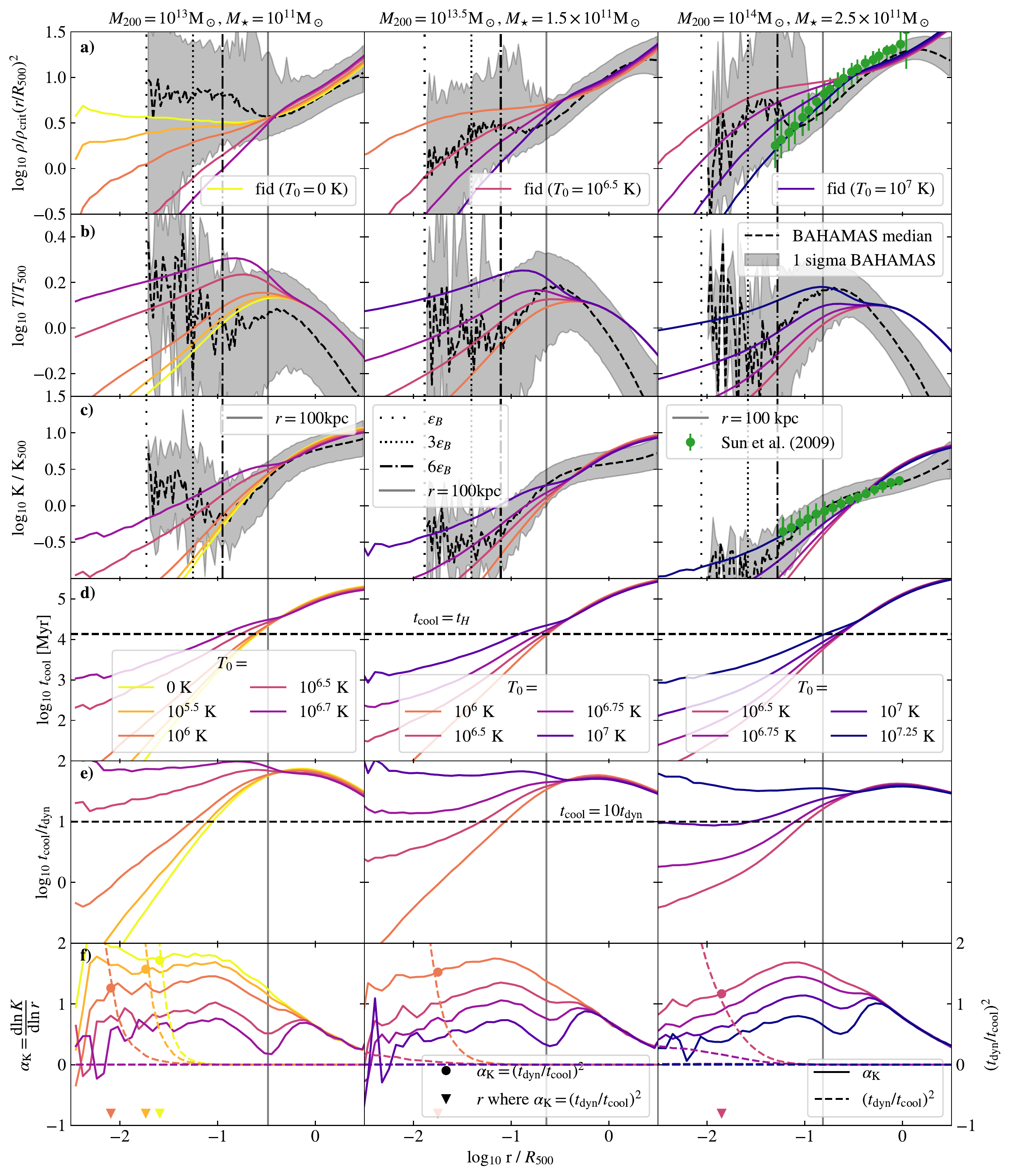}
    \caption{Comparison of the initial median mass-weighted thermodynamic radial profiles to the results from the BAHAMAS cosmological simulation at $z=0$. Different columns correspond to different halo masses (from left to right: $M_{200}/\rm M_\odot = 10^{13}$, $10^{13.5}$ and $10^{14}$). Row (a) shows the density for all gas, row (b) the temperature of hot gas ($T>10^{5.5}~\rm K$), row (c) shows the entropy of the hot gas ($T>10^{5.5}~\rm K$), row (d) the radiative cooling time, row (e) the ratio $t_{\rm cool}/t_{\rm dyn}$, and row (f) the entropy slope as continuous lines and $(t_{\rm dyn}/t_{\rm cool})^2$ as the dashed lines, the point at which the solid and dashed lines intersect is marked by a dot and the corresponding triangle indicates the radius at which this happens. In each column, different colours correspond to different central temperatures. The black dashed line and grey area indicate the mass-weighted median and 1$\sigma$ spread in the BAHAMAS simulation for haloes with the same mass. The vertical grey line indicates $r=100~\rm kpc$. The green points with error bars in the right column show the observations of \citet{sun2009} and the vertical dotted and dash-dotted lines show 1, 3 and 6 times the gravitational softening of the BAHAMAS simulation ($5.7 ~\rm kpc, 17~\rm kpc$ and $34~\rm kpc$). In row (d) the black dashed line indicates the Hubble time ($t_{\rm H}$) and in row (e) the black dashed line shows the value of $t_{\rm cool}/t_{\rm dyn}=10$.}
    \label{fig:BAHAMAS-comparison}
\end{figure*}

\subsection{Calibration against the BAHAMAS simulations}
\label{sec:BAHAMAS}

We use the BAHAMAS simulation to calibrate the two free parameters of our model. For each halo mass the total amount of gas is set by the BAHAMAS gas fraction within $R_{500}$, $f_{500}$ for each halo (see Table \ref{tbl:extra_params}). Determining $T_0$ is less straightforward since it predominantly affects the centre of the halo where the resolution of BAHAMAS becomes a limiting factor. We therefore generated a range of haloes with different values of $T_0$ for each $M_{200}$ and compare these to BAHAMAS in Fig.~\ref{fig:BAHAMAS-comparison}. Specifically, we compare the radial density, temperature, and entropy profiles, where the entropy is defined as $K\equiv k_{\rm B} T / n_{\rm e}^{2/3}$. For clarity, we normalised the temperature and entropy profiles to their analytic virial values within $R_{500}$, namely $T_{500} = GM_{500} \mu m_{\rm p}/2 k_{\rm B} r_{500}$ and $K_{500} = k_{\rm B} T_{500}/[ 500 f_{\rm b} (\rho_{\rm crit}/\mu_{\rm e} m_{\rm p} ) ]^{2/3}$, where $\mu_{\rm e}=1.14$ is the mean atomic weight per free electron and $f_{\rm b}$ the cosmic baryon fraction. To reduce the dynamic range in Fig.~\ref{fig:BAHAMAS-comparison}, the density profile is normalised to $\rho_{\rm crit}$ and multiplied by $(r/R_{500})^2$. The three columns of Fig.~\ref{fig:BAHAMAS-comparison} show the normalised profiles for haloes with $M_{200}=10^{13}$, $10^{13.5}$ and $10^{14}~\rm M_\odot$ respectively, as solid lines coloured according to $T_0$. Black dashed lines show the corresponding mass-weighted median stacked profiles of BAHAMAS haloes with mass within $\Delta \log_{10} M_{500} = 0.01$ at $z=0$, with the 16th and 84th percentiles of the gas particle distribution represented by grey bands. For both the entropy and temperature profiles we only select gas that has a temperature of $T>10^{5.5}~\rm K$ in order to crudely mimic the X-ray observations that were used to calibrate BAHAMAS. For the density profile we select all gas because we want to compare the BAHAMAS total gas density profile\footnote{This means that the profile includes the contribution from satellite galaxies.} with the density profile from the idealised set-up.

Because the density profiles are normalised to $f_{500}$ for BAHAMAS the normalisation agrees for different values of $T_0$, but the shape is different. The temperature and entropy agree with BAHAMAS in the inner region. At large radii there is however a disagreement between the idealised profiles and BAHAMAS, most clearly in the temperature and entropy profiles for higher halo masses. This disagreement is probably because the idealised profiles do not take the cosmological environment into account, namely that gas is accreted and/or a significant amount of gas is in satellite galaxies.  

Keeping these discrepancies in mind, for the $10^{13}~\rm M_\odot$ haloes $T_{0}=0~\rm K$ matches the BAHAMAS profiles most closely. For $M_{200}=10^{13.5}~\rm M_\odot$, $T_{\rm 0} = 10^{6.5}~\rm K$ is the best model. Although $T_{\rm 0} = 10^{6.75}~\rm K$ matches the entropy profile better, it underpredicts the central density by up to a factor 3. For the $10^{14}~\rm M_\odot$ halo the optimal value of $T_{\rm 0}$ is $10^7~\rm K$. It also matches with the observations of \citet{sun2009}. $T_{\rm 0} = 10^{7.25}~\rm K$ matches the entropy better, but has a large cooling time ($t_{\rm cool}>1~\rm Gyr$) making it less ideal to study the BH/ICM connection (see below).

The initial conditions presented in this section are made publicly available as examples distributed with the \texttt{SWIFT} code\footnote{\texttt{swift} is available at \url{http://www.swiftsim.com}.}.

%% file: simulations.tex
\section{Simulations}
\label{sec:subgrid}

Starting from the initial conditions described above, we have run a suite of simulations with the N-body + hydrodynamics code \texttt{SWIFT} \citep{schaller16,schaller2018}. Self-gravity of the baryons is solved with a fast multipole method \citep{greengard1987} while the external gravitational forces from the dark matter halo is approximated by a static external potential (see \S \ref{sec:IC}). Self-gravity is softened with a Plummer-equivalent softening length of $\varepsilon_{\rm grav} = 300~\rm pc$. Particle time steps are fully adaptive, based on the local acceleration ($\Delta t \propto 1/\sqrt{|\mathbf{a}|}$) and limited to 1 per cent of the circular orbital period at the particle's position. For gas, we additionally apply the Courant–Friedrichs–Lewy (CFL) condition and the \citet{durier2012} time step limiter in the vicinity of recent feedback events. For the hydrodynamics, we use SPHENIX \citep{borrow2020}, a density-energy based SPH scheme that includes artificial conduction and artificial viscosity to capture shocks and contact discontinuities. The smoothing lengths of the high-resolution particles are adaptively set to $1.2348$ times the local inter-particle separation (corresponding to 58 neighbours), limited to a minimum of $0.01 \varepsilon_{\rm grav} = 3~\rm pc$. The gas particles have a maximal smoothing length $h_{\rm max}=250~\rm kpc$, and the high-resolution particles all have $h<5~\rm kpc$.

\subsection{Radiative cooling and star formation}
Radiative gas cooling and heating is implemented using the tables of \citet{ploeckinger2020}, which used \texttt{CLOUDY} \citep{ferland2017} to calculate the individual equilibrium cooling and heating rates for the 11 most important elements (H, He, C, N, O, Ne, Mg, Si, S, Ca, Fe), in the presence of a spatially uniform, time-varying UV background based on the model of \citet{faucher-giguere2020}, interstellar radiation and cosmic rays that depend on local gas properties through the Kennicutt-Schmidt (KS) relation \citep{kennicutt1998}, and dust (assuming a constant dust-to-gas ratio of $(D/G)=5.6\times 10^{-3}$ at high densities and a lower dust-to-gas ratio at lower densities that scales also with the KS relation). The cooling tables account for self-shielding of gas, using a cloud size given by the local Jeans length. 

Our simulations do not model the cold phase ($T \la 10^4~\rm K$) of the ISM. We therefore impose a temperature floor corresponding to a constant Jeans mass as proposed by \citet{schaye2008},
\begin{align}
 T_{\rm floor} &= T_{\rm floor, 0} \left(\frac{\rho_{\rm g}}{\rho_{\rm g,0}}\right)^{1/3},
\end{align}
normalised to $T_0=8000~\rm K$ at a density of $n_{\rm H} =0.1~\rm cm^{-3}$. Numerically, this is implemented as an entropy floor and we convert between entropy and temperature assuming a hydrogen mass fraction $X=0.756$ and mean particle mass $\mu=1.22$ of neutral primordial gas for all gas particles.


Gas with entropy within $0.3~\rm dex$ of the floor is assumed to be multi-phase, with an unresolved cold dense component. We use the tables of \citet{ploeckinger2020} to calculate a subgrid temperature $T_{\rm subgrid}$ of this cold phase, and the corresponding subgrid density $\rho_{\rm subgrid}$, by assuming that the gas is in thermal and pressure equilibrium\footnote{For gas that is more than $0.3~\rm dex$ above the floor, the subgrid temperature and density are identical to the SPH temperature and density.}. Gas with $T_{\rm subgrid}< 8\times 10^3~\rm K$ is considered to be star-forming. This mainly selects gas that is on the temperature floor, with a star formation rate based on its pressure following \citet{schaye2008}.

\subsection{Stellar feedback and chemical enrichment}
Once formed, a star particle is assumed to represent a simple stellar population (SSP) with uniform metallicity and formation time, each following a \citet{chabrier2003} initial mass function (IMF) with a mass range of $0.1-100~\rm M_\odot$. For the stellar enrichment we follow \citet{wiersma2009}, with the modifications described in \citet{schaye2015}, which accounts for stellar winds from AGB and massive stars, core collapse supernovae (SNII) and type Ia supernovae (SNIa). 

We assume that massive stars with $M>8~\rm M_\odot$ end their lives as SNII, corresponding to $n_{\rm SNII} = 1.18\times 10^{-2}~\rm M_\odot^{-1}$ SNe II per unit stellar mass formed, each releasing $10^{51}~\rm erg$ of energy. Instead of a fixed SNII delay \citep[as in][]{schaye2015}, we calculate the expected number of SNII during each time step, $N_{\rm SNII}$, based on the \citet{portinari1998} lifetimes of individual stars within the SSP. 
To prevent numerical overcooling, we follow \citet{dalla-vecchia2012} and use the total energy from these SNe to heat a small number of gas particles by $\Delta T=10^{7.5}~\rm K$. Energy feedback from stars is implemented in stochastic, thermal and isotropic form as described in detail by \citet{chaikin2022}\footnote{In essence, we draw a ray in a random direction and select the SPH gas neighbour that has the smallest angular separation from this ray.}. 

Besides SNII, we also implement SNIa feedback. In clusters and their quenched BCGs, SNIa are considered the second most important energy source after AGN and may help quench the galaxy by heating gas at larger radii \citep[e.g.][]{ciotti2007}. In addition, they contribute significantly to the metal enrichment of the ICM \citep[e.g.][]{simionescu2015}. It is therefore important to prevent numerical overcooling of SNIa feedback --- a subtlety not considered in most simulations. We achieve this with the same implementation as described above for SNII feedback. In contrast to SNII, SNIa can occur long after star formation and the precise delay cannot be predicted theoretically because it depends on poorly constrained parameters such as the binary fraction and separation. We therefore take a statistical approach and sample the SNIa rate from a delay time distribution (DTD) in every time step. In general, the DTD can be written as $ \text{DTD}(t) = \nu \xi(t)$, where $\nu$ is the total number of SNIa per unit formed stellar mass and $\xi(t)$ is a function normalised to
\begin{align}
 \int\limits_0^{t_{\rm u}} \xi(t) \mathrm{d}t &= 1, \label{eq:DTDnorm}
\end{align}
where $t_{\rm u}$ is the age of the Universe\footnote{Because most DTDs do not converge up to infinity, it is a convention to integrate up to the age of the Universe, $t_{\rm u}$, instead.} \citep{maoz2012}. Most commonly, the DTD is assumed to follow a power law, i.e. $\xi(t) \propto t^{-\beta}$. We set the slope of the power law to $\beta=1.0$, as inferred from the cosmic SNIa rate both in the field and in clusters \citep[see][for reviews]{maoz2012b, maoz2014}. Furthermore, we assume that SNIa have a fixed minimum delay time of $t_{\rm delay}=40~\rm Myr$ corresponding to the maximum lifetime of stars that explode as SNII. The properly normalised DTD is therefore given by
\begin{equation}
 \text{DTD}(t) = 
    \nu \frac{1}{\ln(t_{\rm u}) - \ln(t_{\rm delay})}
    t^{-1}\Theta(t-t_{\rm delay}).
\end{equation}
In contrast to the power-law slope $\beta$, the DTD normalisation $\nu$ is observed to differ significantly between the field and clusters. For the former, observations agree very well on a value of $\nu = 1.2\times 10^{-3} \; \rm M_\odot^{-1}$\citep{maoz2011, maoz2012, graur2011, perrett2012, frohmaier2019, strolger2020}, whereas a higher value of $\nu = 4\times 10^{-3} \;\rm M_\odot^{-1}$ is preferred by both observations of SNIa in galaxy clusters \citep{maoz2010b, maoz2012b, freundlich2020, strolger2020} and cluster iron abundances \citep{maoz2010}. Because we are simulating massive haloes, we adopt the latter value in this work.

\begin{table*}
\begin{center}
\caption{Simulation parameters. From left to right, we list the dark matter halo mass within apertures $R_{200}$, $R_{500}$, and $R_{2500}$ and the corresponding radii; 
  the initial black hole mass, $M_{\rm BH}$;
  the stellar-to-halo mass ratio $M_{\star}/M_{\rm h}$; 
  stellar scale length $r_\star$; 
  concentration of the dark matter halo, $c$, based on \citet{correa2015};
  gas fraction within $R_{500}$, $f_{500}$;
  virial temperature at $R_{500}$, $k_{\rm B} T_{500}$;
  virial entropy at $R_{500}$, $K_{500}$; 
  the fiducial imposed central initial temperature, $T_{0,\rm fid}$.
  }
\label{tbl:extra_params}
\begin{tabular}{cccclcccccclcl}

\hline
   $\log_{10} M_{200}$ & $\log_{10} M_{500}$ & $\log_{10} M_{2500}$ & $R_{200}$ & $R_{500}$ & $R_{2500}$ & $\log_{10} M_{\rm BH}$ & $M_{\star}/M_{\rm 200}$ & $r_\star$ & $c$ & $f_{\rm gas, 500}$ & $k_{\rm B} T_{500}$ & $\log_{10} K_{500}$ & $T_{0,\rm fid}$\\
                $\rm M_\odot$  & $\rm M_\odot$& $\rm M_\odot$ & $\rm kpc$ & $\rm kpc$ & $\rm kpc$ & $\rm M_\odot$         &     & $\rm kpc$ & &  &            $\rm keV$ & $\rm keV~\rm cm^2$  & $\rm K$ \\
\hline 
   13.0 & 12.89 & 12.62 & 443 & 305 & 143 & 8.4 & 0.01 & 4.5 & 7.2 & 0.016 & 0.34 & 1.78 & 0 \\

   13.5 & 13.37 & 13.09 & 650 & 438 & 205 & 8.6 & 0.005 & 5.0 & 6.4 & 0.037 & 0.7 & 2.09 & $10^{6.5}$\\

   14.0 & 13.88 & 13.56 & 955 & 651 & 294 & 8.8 & 0.0025 & 6.5 & 5.6 & 0.045 & 1.6 & 2.43 & $10^7$
\end{tabular}
\end{center}
\end{table*}

\subsection{Black holes and AGN feedback}
Finally, our simulations model the growth of supermassive BHs and the associated energy feedback (AGN feedback). Since we are simulating an idealised halo, with a single BH, we fix its position explicitly to the centre of the dark matter halo. This is because even at our comparatively high resolution, our simulations do not properly resolve the processes (such as dynamical friction) that determine the motion of the BH \citep[e.g.][]{tremmel2017, bahe2021}. 

Instead of modeling gas accretion with the standard Bondi-Hoyle approach, we use a modification that accounts for supersonic turbulence around the BH. The accretion rate is given by,
\begin{align}
    \dot{m}_{\rm accr, turb} &= 4 \pi \rho G^2 \frac{m_{\rm BH}^2}{c_{\rm s}^3} \left[ \frac{\lambda^2 + \mathscr{M}^2}{(1+\mathscr{M}^2)^4}\right]^{1/2},
\end{align}
with $\lambda= 1.1$ and Mach number $\mathscr{M}=v/c_{\rm s}$, where $v$ and $c_{\rm s}$ are the bulk velocity and (SPH-smoothed) sound speed of the gas surrounding the BH \citep{krumholz2006, ruffert1994}. This expression reduces to the Bondi-Hoyle accretion rate for $\mathscr{M}\ll 1$, but for highly supersonic gas flows around the BH ($\mathscr{M} \gg 1$), the accretion rate is suppressed by a factor $\mathscr{M}^{-3}$. 

When the gas flow around the BH is vorticity-dominated, we instead use the BH accretion rate of \citet{krumholz2005},
\begin{align}
    \dot{m}_{\rm accr, ang} &= \frac{4 \pi \rho (Gm_{\rm BH})^2}{c_{\rm s}^3} 0.34 f(\omega_\star),
\end{align}
where $\omega_\star = \omega r_{\rm B} / c_{\rm s}$, $r_{\rm B} = G m_{\rm BH}/c_{\rm s}^2$ is the Bondi radius, $\omega = |\nabla \times \mathbf{v}|$ is the vorticity of the ambient gas and the function $f(\omega_\star)$ is given by $f(\omega_\star) = (1+\omega_\star^{0.9})^{-1}$. To combine these two approaches, we interpolate the accretion rate between them as
\begin{align}
    \dot{m}_{\rm accr} &= \left( \dot{m}_{\rm accr, turb}^{-2} + \dot{m}_{\rm accr, ang}^{-2} \right)^{-1/2}.
\end{align}
The gas accretion rate is limited to the Eddington rate,
\begin{align}
    \dot{m}_{\rm accr} &= {\rm min}\left(\dot{m}_{\rm accr}, \dot{m}_{\rm Edd} \right),
\end{align}
where
\begin{align}
    \dot{m}_{\rm Edd} &= \frac{4 \pi G m_{\rm BH} m_{\rm p}}{\epsilon_{\rm r} \sigma_{\rm T} c} = 2.218 ~{\rm M_\odot} \; {\rm yr^{-1}} \left( \frac{m_{\rm BH}}{10^8 ~\rm M_\odot} \right).
\end{align}
Here, $\sigma_{\rm T}$ is the Thomson cross-section, $\epsilon_{\rm r}=0.1$ is the assumed radiative efficiency of the accretion disk \citep{shakura1973} and $c$ is the speed of light. 

The mass growth of the BH is then given by
\begin{align}
    \dot{m}_{\rm BH} &= (1 - \epsilon_{\rm r}) \dot{m}_{\rm accr},
\end{align}
while the rest of the accreted mass is radiated away (see below). The mass gain of the BH during a time step $\Delta t$ is therefore
\begin{align}
    \Delta m_{\rm BH} =  (1-\epsilon_r)\dot{m}_{\rm accr} \Delta t.
\end{align}
To conserve the total mass of our simulation (aside from radiative losses), we follow \citet{bahe2021} and remove small amounts of mass from neighbouring gas particles at each time step, as long as this would not reduce the gas particle mass below half their initial mass\footnote{Since the position of the BH is fixed in our simulation, we do not transfer momentum from gas particles to the BH, which leads to an (insignificant) violation of momentum conservation.}.

AGN feedback is implemented based on \citet{booth2009}. The amount of energy available for AGN feedback is given by
\begin{equation}
    \dot{E}_{\rm BH } = \epsilon_{\rm f} \epsilon_{\rm r} \dot{m}_{\rm accr} c^2 = 1.787 \times 10^{56} {~\rm erg ~\rm Myr^{-1}} \left( \cfrac{\dot{m}_{\rm accr}}{10^{-2}~\rm M_\odot ~\rm yr^{-1}} \right), 
\end{equation}
where we assume $\epsilon_{\rm f}=0.1$ for the coupling coefficient of the AGN energy to the ambient gas. In each time step, the energy released to the surrounding gas is stored in an (initially empty) reservoir $E_{\rm BH, res}$ in which this energy is accumulated until it reaches a threshold $E_{\rm heat}$. The latter corresponds to the energy required to heat a single gas particle by $\Delta T=10^{8.5}~\rm K$. Such a high heating temperature is needed to prevent numerical overcooling. At this point we determine the largest integer number $N$ of gas particles that can be heated by $\Delta T$, which may be $>1$ if the BH has accreted substantial mass in the current time step (but see below). The same procedure as for SN feedback is used to select the particles to heat. The energy used to heat them is then subtracted from the energy reservoir.

In order to accurately follow the accretion of gas onto the BH, we want to avoid artificially delaying the onset of AGN feedback through too long time steps. We therefore use an additional time step limiter for the BH to prevent an excessive build-up of energy in its reservoir,
\begin{align}
    \Delta t_{\rm BH, accr} &= \frac{E_{\rm heat}}{\dot{E}_{\rm BH}} =  \frac{k_{\rm B} \Delta T_{\rm AGN} m_{\rm gas}}{(\gamma - 1) \mu m_{\rm H} \epsilon_{\rm f} \epsilon_{\rm r} \dot{m}_{\rm accr} c^2}. \label{eq:timestep1}
\end{align}
In other words, at a constant accretion rate $\dot{m}_{\rm accr}$ the energy reservoir should gain at most $E_{\rm heat}$ (the energy needed to heat one gas particle) during the BH time step. However, in particular for simulations of more massive haloes than presented here, very high accretion rates (and hence $\dot{E}$) could require unacceptably small time step ($t \approx 50~\rm yr$ and $t \approx 10~\rm yr$ for $10^{8.8}~\rm M_\odot$ and $10^{9.5} ~\rm M_\odot$ BHs accreting at the Eddington limit, respectively). To prevent such very small time steps, $\Delta t_{\rm BH, \rm accr}$ is restricted to a minimum of $\Delta t_{\rm BH, \rm accr}=100~\rm yr$.

\subsection{Simulation runs}
For each of the three different halo masses ($10^{13}~\rm M_\odot$, $10^{13.5}~\rm M_\odot$ and $10^{14}~\rm M_\odot$), we run a suite of simulations that keep the parameters listed in Table \ref{tbl:extra_params} fixed, but adopt different values of the initial central temperature $T_0$. For the $10^{13}~\rm M_\odot$ halo we run five simulations ($T_{\rm 0}=\{ 0, 10^{5.5}, 10^6, 10^{6.5}, 10^{6.7}\}~\rm K$, and four each for $10^{13.5}~\rm M_\odot$ ($T_{\rm 0}=\{ 10^{6}, 10^{6.5}, 10^{6.75}, 10^7 \}~\rm K$) and $10^{14}~\rm M_\odot$ ($T_{\rm 0}=\{ 10^{6.5}, 10^{6.75}, 10^{7}, 10^{7.25}\}~\rm K$). To study the long-term evolution of galaxy clusters each simulation is run for $8~\rm Gyr$, which is approximately equal to the time between $z=1$ and $z=0$.

%% file: initial_phase.tex
We begin our analysis by investigating the overall features of the simulated clusters before focusing in detail on the ICM-AGN connection in \S \ref{sec:ICM-AGN}. In particular, we describe the features emerging in the simulated ICM (\S \ref{subsec:spatialscales}, \S \ref{subsec:risebubbles}) and its radially averaged thermodynamical profiles (\S \ref{subsec:thermoprof}) before analysing the evolution of the SFR and its dependence on the parameter $T_0$ (\S \ref{subsec:sensitivity-SFR}).

\begin{figure*}
	\includegraphics[width=.97\linewidth]{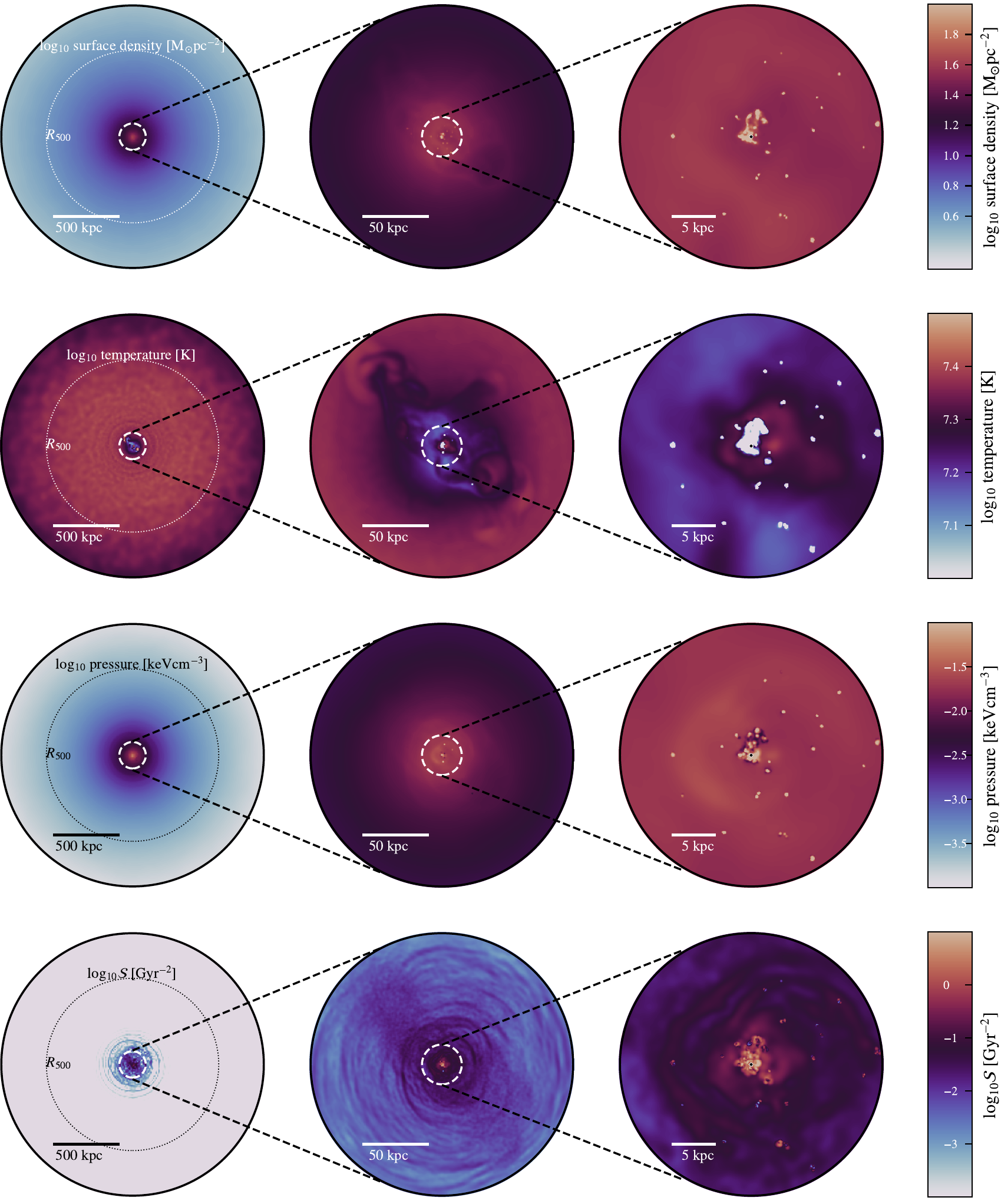}
    \caption{Comparison of the gas surface density (top row, within $\pm 25~\rm kpc$ (left and middle) or $\pm 15~\rm kpc$ (right) along the line of sight), mass-weighted temperature (second row, infinitely thin slice), mass-weighted  pressure (third row, infinitely thin slice) and mass-weighted shock indicator (bottom row, infinitely thin slice, see text for details) at different physical scales for the $M_{200} =10^{14}~{\rm M}_\odot$ halo. The dotted line in the left panels indicates $R_{500}$. In the left and middle panels a dashed white line indicates the extent of the image to its right. The physical scale is indicated with the black or white bar. In the right columns, black dots within a white circle indicate the position of the BH. All images correspond to time $t=3~\rm Gyr$ from the start of the simulation. The image projection is taken in the plane of the rotation (i.e. the angular momentum vector is pointing out of the paper). Note that some small-scale gas clumps have temperatures and densities well below and above our adopted scaling range, respectively. }
    \label{fig:phases}
\end{figure*}

\begin{figure}
    \centering
	\includegraphics[width=.7\columnwidth]{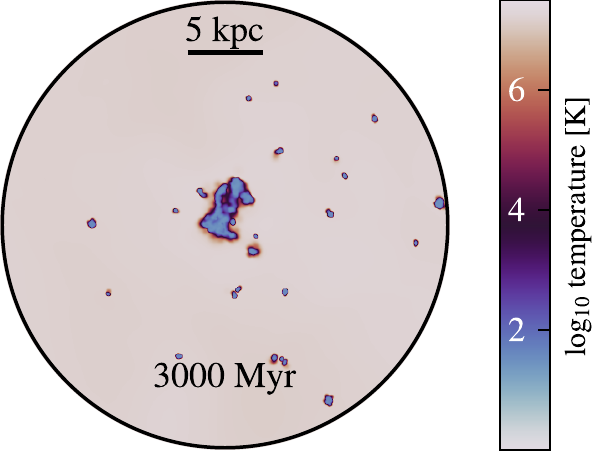}
    \caption{Mass-weighted logarithmic-averaged temperature distribution in the inner $15~\rm kpc$ of the $M_{200}=10^{14}~\rm M_\odot$ halo. The cold gas in the centre of the halo is similar to structures observed in H$\alpha$ emission.}
    \label{fig:clumpy-temperature}
\end{figure}

\begin{figure*}
	\includegraphics[width=.98\linewidth]{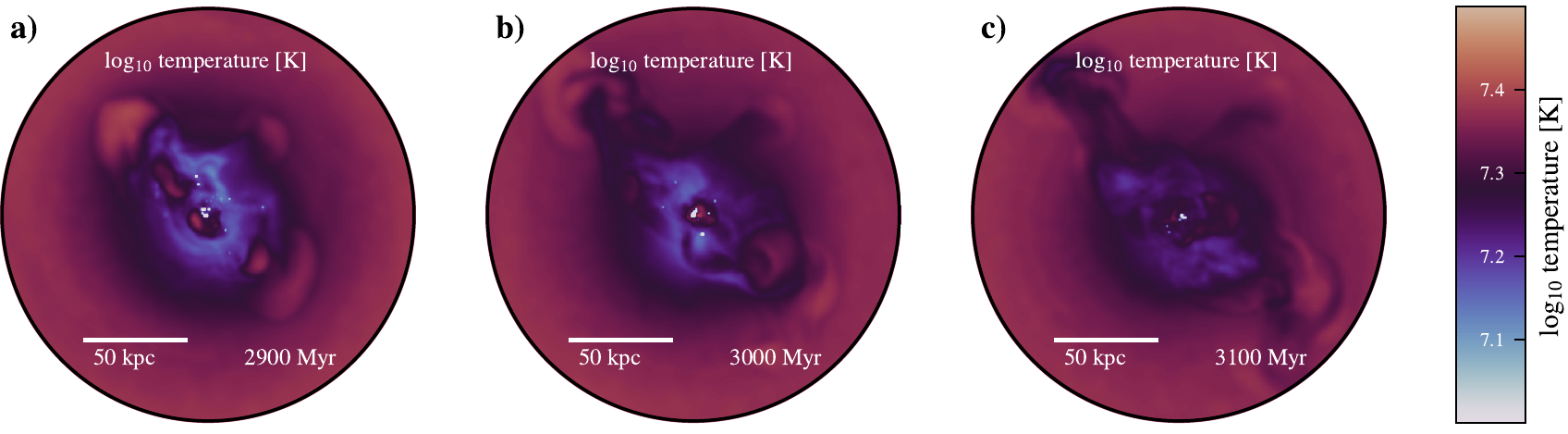}
    \caption{Projected gas temperature of the fiducial $10^{14}~\rm M_\odot$ halo in infinitely thin slices at three times separated by $100~\rm Myr$ each ($t=$ 2900 Myr, 3000 Myr and 3100 Myr). The first time frame ($t=2900~\rm Myr$) corresponds to a peak in the SFR and BHAR. The line of sight is approximately perpendicular to the motion of the two prominent bubbles. The online supplementary material shows the evolution more clearly with a movie. Despite a purely isotropic implementation of AGN feedback, the interaction of outflows with the ISM and ICM produces a clearly biconical structure with prominent temperature variations at fixed radius.  
    }
    \label{fig:phases-temp}
\end{figure*}

\subsection{The cluster at different spatial scales} \label{subsec:spatialscales}
To give an impression of how different thermodynamic properties vary on different spatial scales, we first inspect the simulation at a fixed time of $3~\rm Gyr$ for the fiducial simulation of the $M_{200} = 10^{14}~\rm M_\odot$ halo with $T_{\rm min}=10^7~\rm K$. Fig.~\ref{fig:phases} shows the gas mass surface density, temperature, pressure and time derivative of the velocity divergence (defined below) in its four rows. The three columns show images at different spatial scales, with diameters of $2~\rm Mpc$, $200~\rm kpc$ and $30~\rm kpc$, the middle column correspond to our high-resolution region. The online supplementary material has videos that show the evolution of the $200~\rm kpc$ spatial scale for each fiducial halo. 

The central density image reveals the existence of $\approx 20~\rm kpc$ regions with lower density (top middle panel), indicative of previous AGN feedback. These low-density regions have higher-than-average temperatures (middle panel of the 2\textsuperscript{nd} row), such that they are in pressure pressure equilibrium (absence of features in the 3\textsuperscript{rd} row). 

The central $\approx 10~\rm kpc$ of the halo contain many high-density clouds and filamentary structures. These form due to large-scale in-situ cooling which produces large clumps that start collapsing and correspondingly lose their pressure support. This causes the clumps to fall into the centre while rotating around the BH. Close to the BH the large clumps of cold gas are shattered and smaller clumps are brought to larger radii due to the hot bubbles produced by AGN feedback. As seen in Fig.~\ref{fig:clumpy-temperature}, the temperature in the clouds is $< 10^4~\rm K$, indicative of ISM that would be observable in e.g. H$\alpha$ or CO emission (e.g. \citealt{olivares2019}; \citealt{russell2019}). Returning to Fig.~\ref{fig:phases}, we note that clouds have higher pressures than their surrounding ICM (indicated by the light yellow, third row, right panel) implying that they are gravitationally bound. We will return to the importance of these cold gas clouds for the ICM-AGN interaction in \S \ref{sec:ICM-AGN}. 

Shocks and sound waves in the ICM provide an observable record of past AGN episodes that complement the bubbles discussed above \citep[see e.g.][]{sanders2016}. The bottom row of Fig.~\ref{fig:phases} shows a shock/sound wave tracer defined as in \citet{borrow2020}:
\begin{equation}
    \mathcal{S} = \left\{
  \begin{array}{ll}
    \left| \frac{\rm d}{{\rm d} t} \nabla \cdot \mathbf{v}\right| & :  \frac{\rm d}{{\rm d} t} \nabla \cdot \mathbf{v} < 0 ~{\rm or} ~ \nabla \cdot \mathbf{v} < 0,\\
    0 & : \rm otherwise.
  \end{array}
\right.
\end{equation}
The high values of $\mathcal{S}$ in the central few kpc are due to ongoing energy injection by the AGN. At radii $\approx 5~\rm kpc$, the values of $\mathcal{S}$ are overall lower, but show clear ring-like features that indicate outflows from previous AGN episodes. As seen in the third row, the pressure is elevated around the edge of this zone, consistent with an AGN-driven shock front. At larger scales ($10-100~\rm kpc$, middle panel of the bottom row) many more of these rings can be seen, tracing successively older AGN feedback episodes. We note that the top left rising bubble produces a signal on its own that makes the wave tracer less clear in this region. As is evident from the bottom row, the waves disappear beyond $\sim 100~\rm kpc$ due to the degrading resolution (see \S \ref{subsubsec:degrading}). By running a uniform-resolution simulation for $4~\rm Gyr$ we have verified that waves extend all the way to the edge of our simulated halo.

\subsection{The rise of hot buoyant bubbles} \label{subsec:risebubbles}

To investigate the formation of the biconical bubble structure that can be seen in the temperature distribution at radii of $\approx 100~\rm kpc$, Fig.~\ref{fig:phases-temp} shows the temperature map in three consecutive time frames separated by $100~\rm Myr$; the central image (b) corresponds to the time shown in Fig.~\ref{fig:phases}. The left-most image (a) is just slightly after a peak in the SFR and BHAR (see Fig.~\ref{fig:SFR-BHAR-corr} below). Panel (a) shows two large bubbles at an altitude of around $50~\rm kpc$, formed by the resulting AGN feedback, and several smaller bubbles at smaller radii (some of which are in fact moving in a different direction). The online supplementary material shows the time evolution for all three fiducial haloes. The large bubbles move outwards (central panel), in a way reminiscent of an atomic mushroom cloud, in which they push away the gas above them and produce an inflow of colder gas from lower altitudes. The afterwinds of the hot rising bubbles draw in gas that is colder by at least $0.15$ dex, and lift it to altitudes as high as $100~\rm kpc$ (panel c). This demonstrates that AGN feedback is able to produce fluctuations in the ICM at large distances from the BH. In panel (c) the bubble has risen to its maximum altitude, after which the hotter and colder gas start to disperse and the bubble signature fades away (not shown) over the next $400~\rm Myr$.

We remind the reader that this biconical outflow emerges even though we do not use an explicitly biconical model for AGN feedback. The interaction of the intrinsically isotropic AGN feedback with the inhomogeneous ISM and ICM naturally results in the development of biconical structures in a fully self-consistent fashion. This is consistent with what is predicted by the (cosmological) RomulusC simulation \citep{tremmel2019} and highlights that outflow morphologies on scales of tens of kpc are not necessarily indicative of processes occurring close to the BH.

\begin{figure*}
	\includegraphics[width=.98\linewidth]{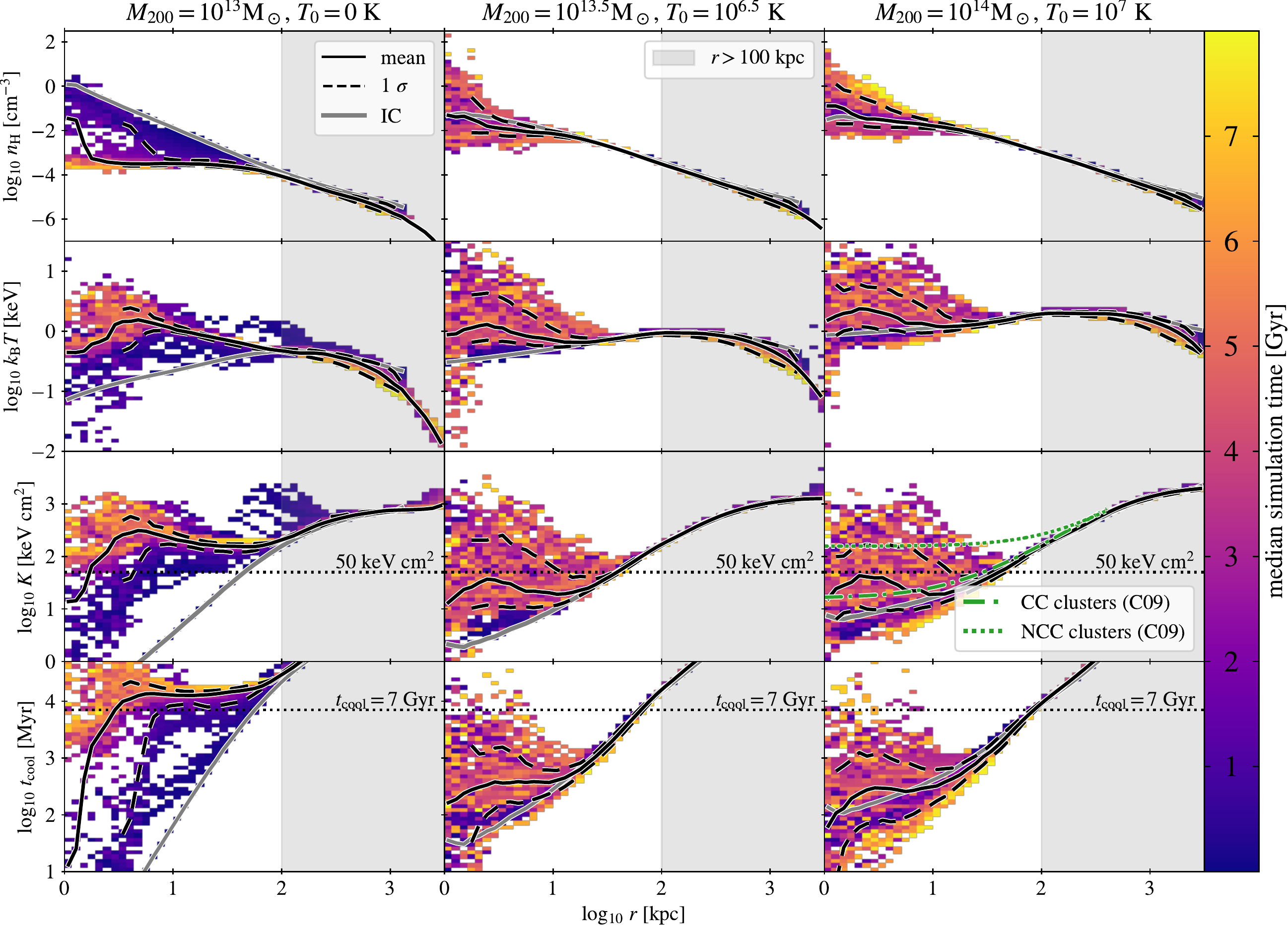}
    \caption{Radially-averaged density (top row), temperature (second row), entropy profiles (third row), and cooling time (bottom row) for the three fiducial simulations of halo masses $M_{200}=10^{13}~\rm M_\odot$, $10^{13.5}~\rm M_\odot$ and $10^{14}~\rm M_\odot$ (left, middle, and right column, respectively). The solid black lines show the volume-weighted median profile based on 160 snapshots spanning 8 Gyr; black dashed lines give the corresponding 16\textsuperscript{th} and 84\textsuperscript{th} percentiles, the grey lines correspond to the initial conditions. Background colours indicate the median time that a particular entropy was reached at each radius. The grey region corresponds to $r>100~\rm kpc$. In the third-right panel, we compare our simulations with the observed ACCEPT galaxy cluster sample \citep{cavagnolo2009}, split into CC and NCC by a central entropy of $50~\rm keV ~\rm cm^2$. The low-mass halo ($M_{200}=10^{13}~\rm M_\odot$) is quickly converted from a CC to an NCC profile. While there is significant scatter in the inner $30~\rm kpc$ with values between CC and NCC, at large radii the scatter is negligible. For the massive haloes $t_{\rm cool}$ is always $\la 1~\rm Gyr$ for $r\la 20~\rm kpc$. $t_{\rm cool}$ decreases with time for gas at $r \ga 10~\rm kpc$, this means that it is inevitable to have new episodes of gas cooling that bring cold gas to the centre. 
    }
    \label{fig:entropy-mean}
\end{figure*}

\begin{figure*}
	\includegraphics[width=.97\linewidth]{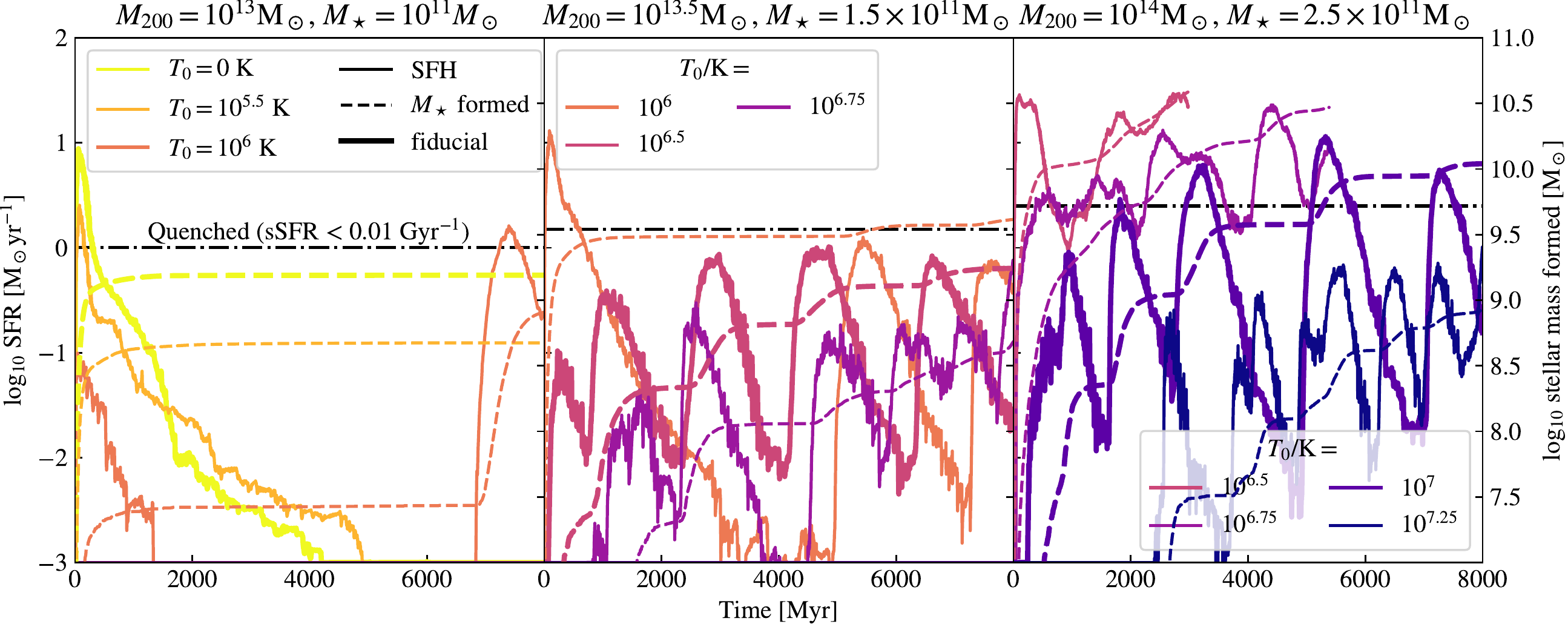}
    \caption{Comparison of the star formation history (SFH) for simulations with different initial central minimum temperatures (different colours and thick curves for the fiducial models) for halo masses of $M_{200}=10^{13}~\rm M_\odot$, $10^{13.5}~\rm M_\odot$ and $10^{14}~\rm M_\odot$ (left, middle and right panels, respectively). Solid curves show the instantaneous SFR (left y-axis), while the dashed curves track the cumulative stellar mass formed (right y-axis). The black dash-dotted line indicates the quenching threshold of sSFR $=10^{-2}~\rm Gyr^{-1}$ (see text for details). For the low-mass halo of $M_{200}=10^{13}~\rm M_\odot$, all initial temperatures lead to rapid ($t\la 1~\rm Gyr$) quenching. More massive haloes instead show oscillatory behaviour with cooler (hotter) cores leading to larger (smaller) mean star formation rates. 
    }
    \label{fig:SFH-summary}
\end{figure*}

\subsection{Mean thermodynamic profiles} \label{subsec:thermoprof}
To understand the short-term fluctuations and the long-term evolution of the ICM in a quantitative way, we investigate the thermodynamic ICM profiles. Fig.~\ref{fig:entropy-mean} shows the volume-weighted, radially-averaged median density, temperature, entropy, and cooling time profiles and their scatter based on 160 snapshots spanning $8~\rm Gyr$ for the fiducial simulations of our three halo masses (different columns). We define the cooling time as
\begin{equation}
 t_\text{cool} = \frac{\frac{3}{2} n k_{\rm B} T}{|\Lambda_\text{net}|},
\end{equation}
where $n$ is the gas number density and $\Lambda_{\rm net}$ is the net radiative cooling rate. To construct the profiles, we first compute the volume-weighted median of the particles in radial bins for each halo individually, and then compute the median profile (black solid line, with 16\textsuperscript{th}/84\textsuperscript{th} percentiles of the profiles indicated by dashed lines). The online supplementary material shows videos of the evolution of the radial profiles for all three haloes that show the evolution of the profiles compared to the SFH. 

As is evident from the third row, all haloes initially have a CC entropy profile (grey line), with central values well below the commonly used demarcation line of $50~\rm keV ~\rm cm^{2}$. For the $M_{200}=10^{13}~\rm M_\odot$ halo, AGN feedback subsequently converts this into an NCC entropy profile, with $K \gg 100~\rm keV ~\rm cm^2$ and $t_{\rm cool}>7~\rm Gyr$ down to $\approx 3~\rm kpc$. Correspondingly, the central temperature is increased by an order of magnitude, while the density is decreased by more than a factor of $10^3$. In contrast, both higher-mass haloes retain a CC, with a time-averaged central entropy below $50~\rm keV ~\rm cm^2$, albeit with significant scatter that regularly results in an NCC-like entropy profile. The $t_{\rm cool}$ is regulates to a constant cooling time for $r<20~\rm kpc$. At large radii ($r \ga 30~\rm kpc$, still well below the edge of the high-resolution region), the profiles remain nearly constant over time, plausibly because outflows produced by AGN rarely reach such large radii (except in the low-mass halo at early times), and if they do, they only cover a small fraction of the total volume (see Fig.~\ref{fig:phases-temp}). The temperature profile at $r>100~\rm kpc$ decreases slightly with time, due to a combination of lower resolution and cooling that is not compensated by shock heating from cosmological accretion. 

In the third-right panel of Fig.~\ref{fig:entropy-mean}, we compare these predictions to the observed CC and NCC entropy profiles of the ACCEPT sample of galaxy clusters\footnote{Consistent with our classification, \citet{cavagnolo2009} used a central entropy of $50~\rm keV~\rm cm^2$ to separate CC from NCC clusters.} \citep[green lines,][]{cavagnolo2009}, which are comparable to our most massive halo (typical $T_X \ga 1~\rm keV$). In general, the entropy profile of our simulated halo agrees well with the observed CC clusters, albeit with a slightly higher entropy in the central $\approx 10~\rm kpc$. The upper range of entropies is comparable to the ACCEPT NCC profile. As we discuss in more detail in \S \ref{sec:ICM-AGN}, this suggests a periodic cycling between CC and NCC states for the $10^{14}~\rm M_\odot$ (and $10^{13.5}~\rm M_\odot$) haloes. 

\subsection{Sensitivity of the SFR to the initial central temperature} \label{subsec:sensitivity-SFR}
We now look at the dependence of the SFR on the initial central minimum temperature of the gas profile in the three different mass galaxy clusters.  

\subsubsection{Star formation histories for different initial central temperature} \label{sec:SFH-T0}
Fig.~\ref{fig:SFH-summary} shows the SFH for our simulations, grouped by halo mass (from left to right: $10^{13}$, $10^{13.5}$ and $10^{14}~\rm M_\odot$). Different values of initial central temperature $T_0$ are represented by different colours. For reference, we indicate the commonly adopted criterion for a galaxy to be quenched, $\rm sSFR < 0.01~\rm Gyr^{-1}$ \citep[e.g.][]{weinmann2006}. For this, we convert from sSFR to SFR based on the initial stellar mass of each simulation, the total stellar mass formed during the $8~\rm Gyr$ (dashed lines) is always negligible compared to the initial stellar mass. 

All galaxies with $M_{200} = 10^{13}\,\rm M_\odot$ are quenched quickly ($t\leq 1~\rm Gyr$) and remain quenched at almost all later times\footnote{This also holds for slightly different quenching criteria.}. This implies that AGN feedback is ubiquitously able to quench group-size haloes, independently of the initial temperature profile. The haloes with large $T_0$ (i.e. $T\geq 10^{6.5}\,\rm K$, not shown) do not form stars because they have $t_{\rm cool} \ga 10 \, t_{\rm dyn}$. These haloes do not cool much and their radiative losses are compensated by adiabatic compression that re-heats the gas. 

Higher halo masses display qualitatively different behaviour. The $10^{13.5}~\rm M_\odot$ simulations with higher $T_{\rm 0}$ ($10^{6.5}~\rm K$ and $10^{6.75}~\rm K$) have oscillatory SFHs, with peak SFRs that tend to increase with decreasing $T_{\rm 0}$. This implies that AGN feedback does not erase the initial conditions and therefore the properties of the haloes with different $T_{\rm 0}$ remain distinct. An exception is the lowest-$T_0$ halo ($10^6~\rm K$, orange line) which starts with a strong starburst but is subsequently transformed due to AGN feedback, resulting in SFRs similar to the $10^{6.5}~\rm K$ halo. The halo with $T_0=10^7~\rm K$ is unable to form stars because of its large $t_{\rm cool}$, similar to the high $T_0$ profiles for the $10^{13}~\rm M_\odot$ haloes. 

At even higher halo masses ($10^{14}~\rm M_\odot$), all four simulations retain distinct SFHs throughout, with a strong dependence of SFR on $T_0$. AGN feedback alone is unable to convert haloes with low initial $T_0$ into something resembling those with high initial $T_0$ and low SFR. This indicates that in order to quench SFR in massive haloes with low central temperatures, we need additional processes beyond what is included in these simulations.  In \S \ref{subsec:regul} we will investigate how the oscillatory behaviour in our simulations is regulated.

\subsubsection{Dependence of the initial SFR on the initial central temperature}
The bottom three rows of Fig.~\ref{fig:BAHAMAS-comparison} show the cooling time, the ratio of cooling to dynamical time ($t_{\rm cool}/t_{\rm dyn}$), and the logarithmic entropy slope for the three different halo masses at $t=0~\rm Gyr$. The dynamical time is defined as
\begin{equation}
 t_\text{dyn} =\sqrt{\frac{2 r}{g}}=\sqrt{ \frac{2 r^3}{G M_{\rm encl}(r)} },
\end{equation}
where $g$ is the local gravitational acceleration and $M_{\rm encl}(r)$ is the total enclosed mass at radius $r$. Since $M_{\rm encl}(r)$ is dominated by the stars of the BCG in the centre, and by the dark matter profile at large radii, at a given radius $t_\text{dyn}$ will remain nearly constant during the simulation. 

As expected, a higher $T_{0}$ leads to a longer initial cooling time and a higher ratio $t_{\rm cool} / t_{\rm dyn}$. When $t_{\rm cool} \ga 10\, t_{\rm dyn}$ gas will not start condensing because it cools slowly and the radiative losses can be compensated by adiabatic compression (see \S \ref{sec:SFH-T0}). However, in the runs with $T_0$ resulting in $t_{\rm cool} / t_{\rm dyn} \la 10$, the cooling is faster and adiabatic compression is unable to compensate the cooling. Therefore, the cooling time $t_{\rm cool}$ sets the time scale for gas to cool down and start condensing. Fig.~\ref{fig:SFH-summary} shows that the cooling time in the inner $10~\rm kpc$ depends on the initial $T_0$ so different times for the onset of star formation are expected.

\begin{figure}
	\includegraphics[width=.97\linewidth]{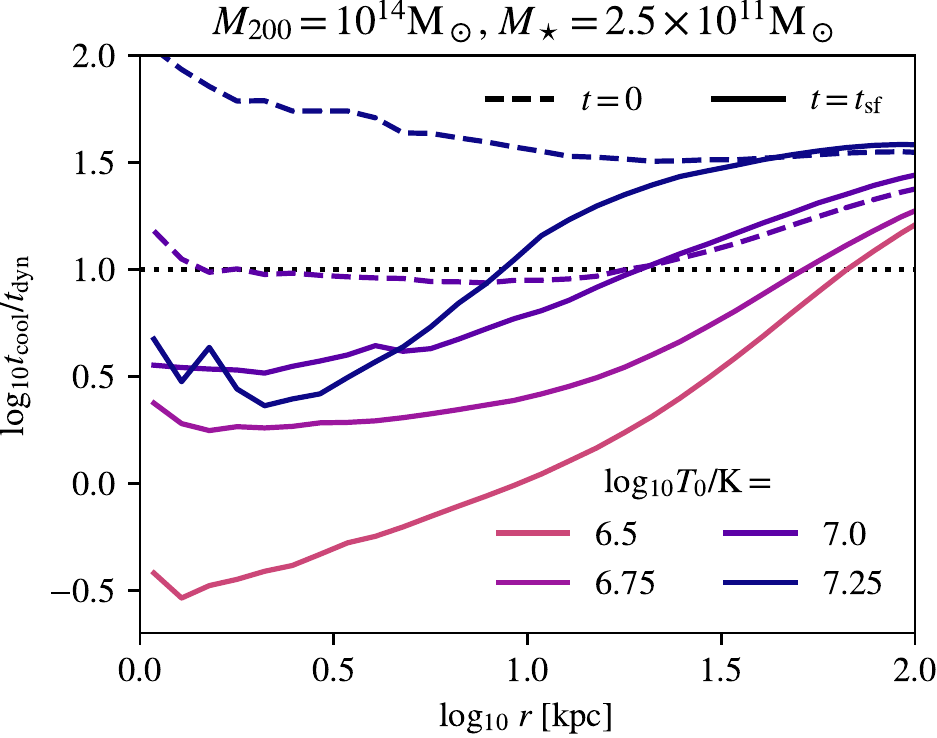}
    \caption{Comparison of $t_{\rm cool}/t_{\rm dyn}$ for the $10^{14}~\rm M_\odot$ halo for different $T_0$ (colours) at time $t=0$ (dashed) and $t=t_{\rm sf}$, immediately before the SFR first reaches $10^{-3}~\rm M_\odot ~\rm yr^{-1}$ (solid line); note that the latter happens already at $t=0$ for $T_0=10^{6.5}~\rm K$ and $10^{6.75}~\rm K$. The black dotted line shows $t_{\rm cool}/t_{\rm dyn}=10$, the threshold for gas to precipitate. When star formation stars, $t_{\rm cool}/t_{\rm dyn} \ll 10$ in the inner $\approx 10-50 ~\rm kpc$, indicating that gas precipitation is not the limiting factor for the onset of star formation.
    }
    \label{fig:tcool-tdyn-M14}
\end{figure}

\begin{figure*}
	\includegraphics[width=.97\linewidth]{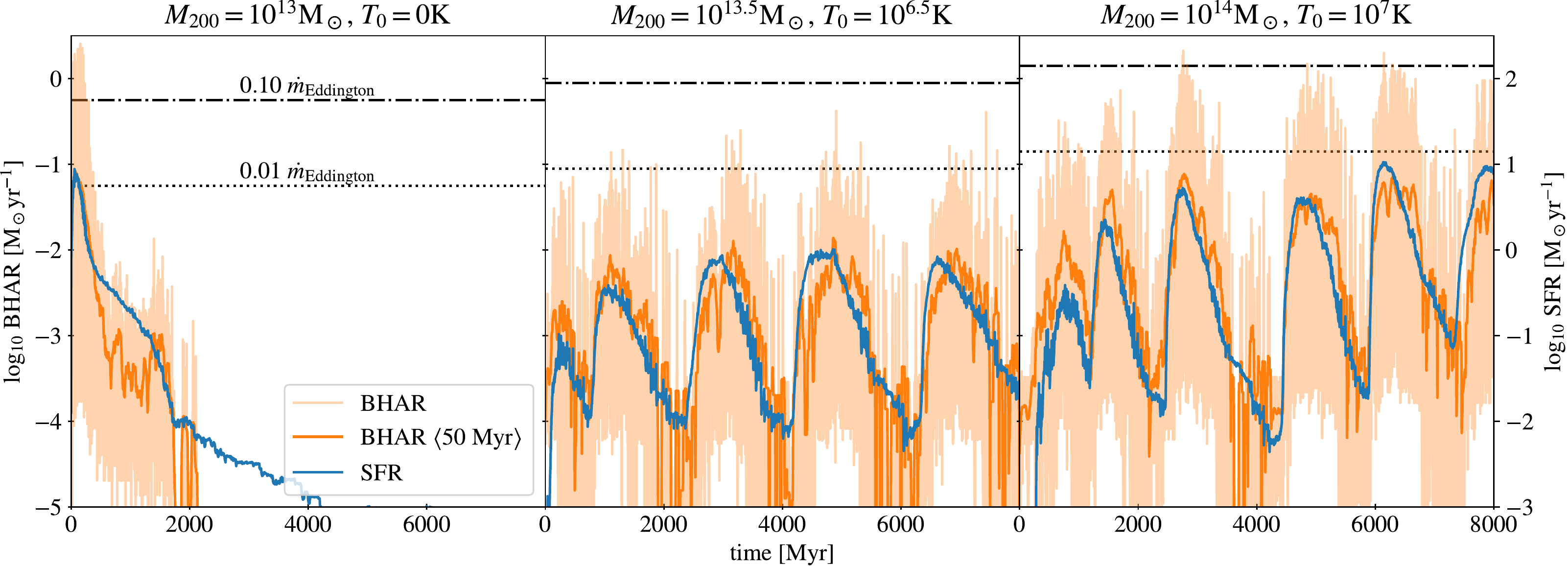}
    \caption{Comparison of the BHAR and SFR for the different fiducial haloes at the three different halo masses (left $M_{200} = 10^{13}~\rm M_\odot$, centre $10^{13.5}~\rm M_\odot$ and right $10^{14}~\rm M_\odot$). The left axis shows the SFR, which is coloured blue in all panels. In each panel the dashed horizontal line indicates a quenching criterion of ${\rm sSFR} < 0.01~\rm Gyr^{-1}$. The right axis shows the BHAR, the grey line shows the BHAR at very high time resolution and the orange line shows the BHAR averaged over $50~\rm Myr$. The dotted and dash-dotted lines show 1\% and 10\% of the Eddington accretion rate, respectively. The instantaneous BHAR varies over many orders of magnitudes over short times and is uncorrelated with the SFR. The 50 Myr averaged BHAR is correlated with the SFR.}
    \label{fig:SFR-BHAR}
\end{figure*}

In order to determine the link between $t_{\rm cool}/t_{\rm dyn}$ and the onset of star formation, we show in Fig.~\ref{fig:tcool-tdyn-M14} the radial profile of $t_{\rm cool}/t_{\rm dyn}$, for the four $T_0$ variants of the $10^{14}~\rm M_\odot$ halo at $t=0$ and for the snapshot at $t_{\rm sf}$, the first time the SFR exceeds $10^{-3}~\rm M_\odot ~\rm yr^{-1}$. At the onset of star formation the central $t_{\rm cool}/t_{\rm dyn}$ has already dropped far below $10$, which indicates that this is not the limiting factor that determines when star formation starts.


\citet{voit2017} derived another criterion to determine whether gas is unstable against condensation and precipitation to the centre of the halo: entropy perturbations are unstable when the logarithmic entropy slope satisfies
\begin{align}
    \alpha_{\rm K} \equiv \frac{{\rm d} \ln K }{ {\rm d} \ln r} \la \left( \frac{t_{\rm cool}}{t_{\rm dyn}} \right)^{-2}. \label{eq:voit}
\end{align}
From the bottom row of Fig.~\ref{fig:BAHAMAS-comparison} and Fig.~\ref{fig:SFH-summary} it is evident that the correlation of this ratio with SFR is less strong than is the case for $t_{\rm cool} / t_{\rm dyn}$. For both the $M_{200}=10^{13.5}~\rm M_\odot$- $T_0=10^{6.5}~\rm K$ and $M_{200}=10^{14}~\rm M_\odot$-$T_0=10^{6.75}~\rm K$ haloes, the \citet{voit2017} criterion does not predict condensation while the SFH (Fig.~\ref{fig:SFH-summary}) shows that in both simulations considerable star formation begins almost immediately after $t=0$. All of the above suggest that the initial cooling time set by the initial temperature profile (i.e. $T_0$) is the key factor that decides when significant star formation begins in our simulations.

%% file: regulation.tex
\subsection{The ICM-AGN connection} \label{sec:ICM-AGN}

The presence of feedback-induced perturbations in the ICM and the periodicity in the SFR that we have seen above are strongly suggestive of a close interplay between the ICM and the central AGN. We now explore this connection.

\subsubsection{How are star formation and black hole accretion regulated?}
\label{subsec:regul}
As a first step, we show in Fig.~\ref{fig:SFR-BHAR} the evolution of the SFR and the BH accretion rate (BHAR) for our fiducial simulations of the three different halo masses. The light orange curves show the instantaneous BHAR, i.e. computed at each simulation time step. Consistent with previous studies of e.g. the cosmological hydrodynamical EAGLE simulation \citep[][]{mcalpine2017} and of idealised galaxy clusters \citep{li2015, qiu2019}, this shows very strong (orders-of-magnitude) fluctuations on $\la ~\rm Myr$ timescales. When averaged over $50~\rm Myr$, however, slower variations become apparent (dark orange lines): a gradual decrease of the time-averaged BHAR down to almost zero for the $10^{13}~\rm M_\odot$ halo, and oscillations with a period of $\approx 2~\rm Gyr$ for the more massive haloes. The (instantaneous) SFR (blue line) correlates strongly with this time-averaged BHAR, but not with the high-frequency fluctuations. 

The oscillations in the SFR and the BHAR are not perfectly synchronized, however: SFR leads by $\sim 10^2~\rm Myr$. Both the strong correlation and the delay are physically expected: both SFR and BHAR depend on the presence of cold dense gas near the centre of the halo, but they are not co-located. After condensation, cold gas can form stars nearly immediately, but it takes more time for it to flow to the BH at the very centre of the halo. This implies an expected delay of 
\begin{align}
    \Delta t \sim 98 ~{\rm Myr} \left( \frac{r}{10~\rm kpc} \right) \left( \frac{v}{100~\rm km ~\rm s^{-1}} \right)^{-1},
\end{align}
where $r$ is the radius within which gas forms stars ($\sim 10~\rm kpc$) and $v$ its radial velocity ($\sim 10^2~\rm km ~\rm s^{-1}$). 

\begin{figure}
	\includegraphics[width=\linewidth]{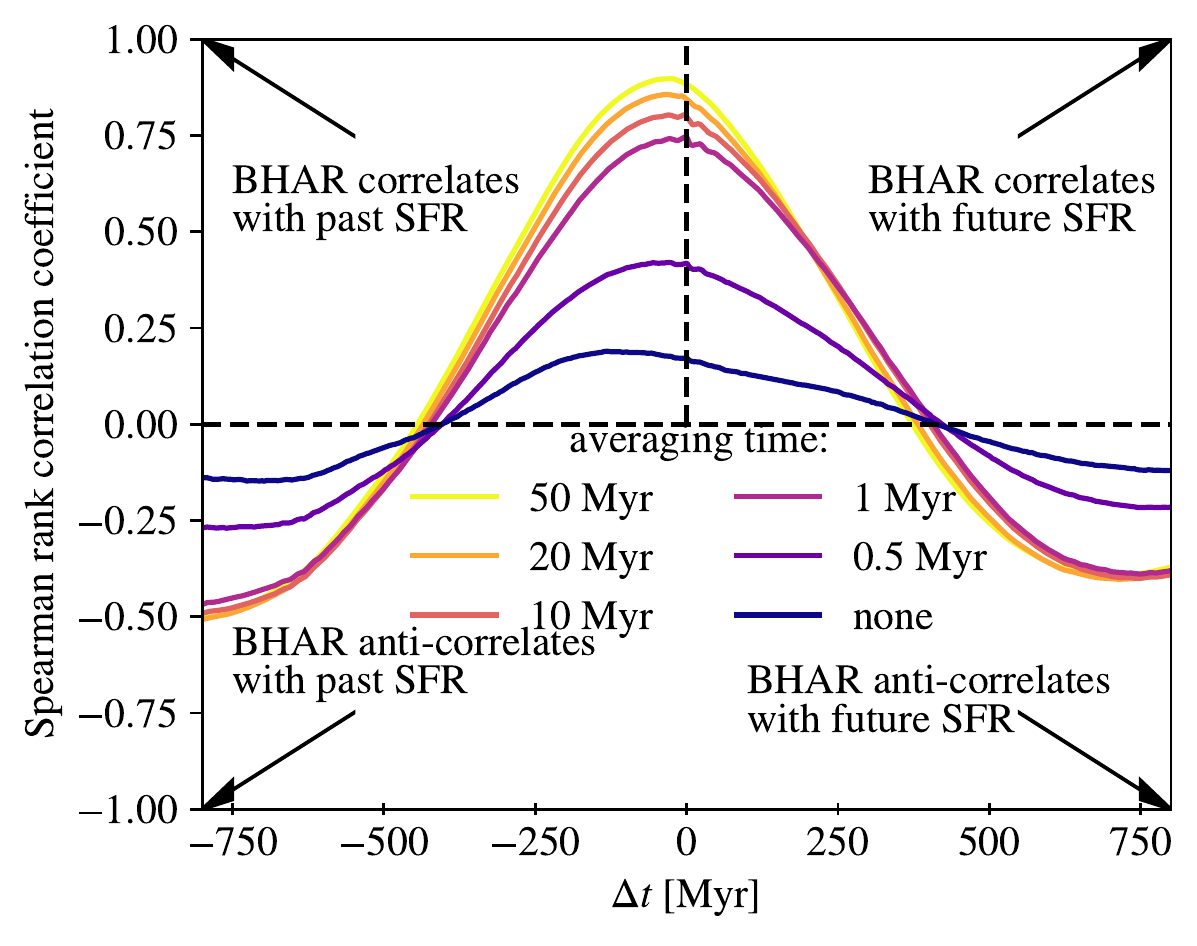}
    \caption{Comparison of the Spearman rank correlation coefficient of SFR and BHAR as a function of delay time $\Delta t$ for different averaging timescales (different colours) for the $M_{200}=10^{14}~\rm M_\odot$ halo. $\Delta t$ is the delay of BHAR with respect to the SFR. The corresponding correlation is calculated as $r_s\left[{\rm BHAR}(t-\Delta t),{\rm SFR}(t) \right]$. We show different values for the time over which the BHAR is averaged (50 Myr to none). When the BHAR is averaged over timescales $t>1~\rm Myr$ there is a strong correlation between the BHAR and SFR, where the past SFR correlates with the current BHAR. There is only a weak correlation when the BHAR is not time-averaged.}
    \label{fig:SFR-BHAR-corr}
\end{figure}

To quantify the offset and correlation between the SFR and the time-averaged BHAR, we use a time-shifted correlation approach. For a given delay $\Delta t$, we calculate the Spearman rank correlation coefficient of ${\rm BHAR}(t-\Delta t)$ and ${\rm SFR}(t)$, i.e. the correlation strength with the BHAR shifted forward (or equivalently the SFR shifted backward) by $\Delta t$. This approach is non-parametric, so that we can remain agnostic about the exact functional form of ${\rm BHAR}(t)$ and ${\rm SFR}(t)$. Fig.~\ref{fig:SFR-BHAR-corr} shows the resulting Spearman rank correlation coefficient for the $10^{14}~\rm M_\odot$ halo. For averaging timescales $t_{\rm avg} \geq 1~\rm Myr$ there is a very strong peak correlation ($r_{\rm s}\approx 0.8$) between the SFR and BHAR at $\Delta t \approx -50~\rm Myr$,  which is almost independent of the precise value of $t_{\rm avg}$. In other words, the peak (and minimum) SFR is followed by the peak (minimum) BHAR after a delay of $\approx 50~\rm Myr$, in good agreement with our analytic estimate. Results for the $10^{13.5}~\rm M_\odot$ are similar (not shown).

For shorter averaging timescales (i.e. $t_{\rm avg}\ll 10~\rm Myr$) the correlation is significantly weaker, suggesting that BHAR variations on these short timescales are not driven by the global availability of cold gas. Instead, they are plausibly caused by the (stochastic) motion of dense clumps relative to the BH, which do not affect the SFR. This means that it may be challenging to detect correlations between SFR and observational tracers of the (near-)instantaneous BHAR, such as the X-ray luminosity of AGNs. Observables sensitive to the energy injected by AGN over longer timescales, such as radio bubbles, are more promising.

\begin{figure*}
	\includegraphics[width=.97\linewidth]{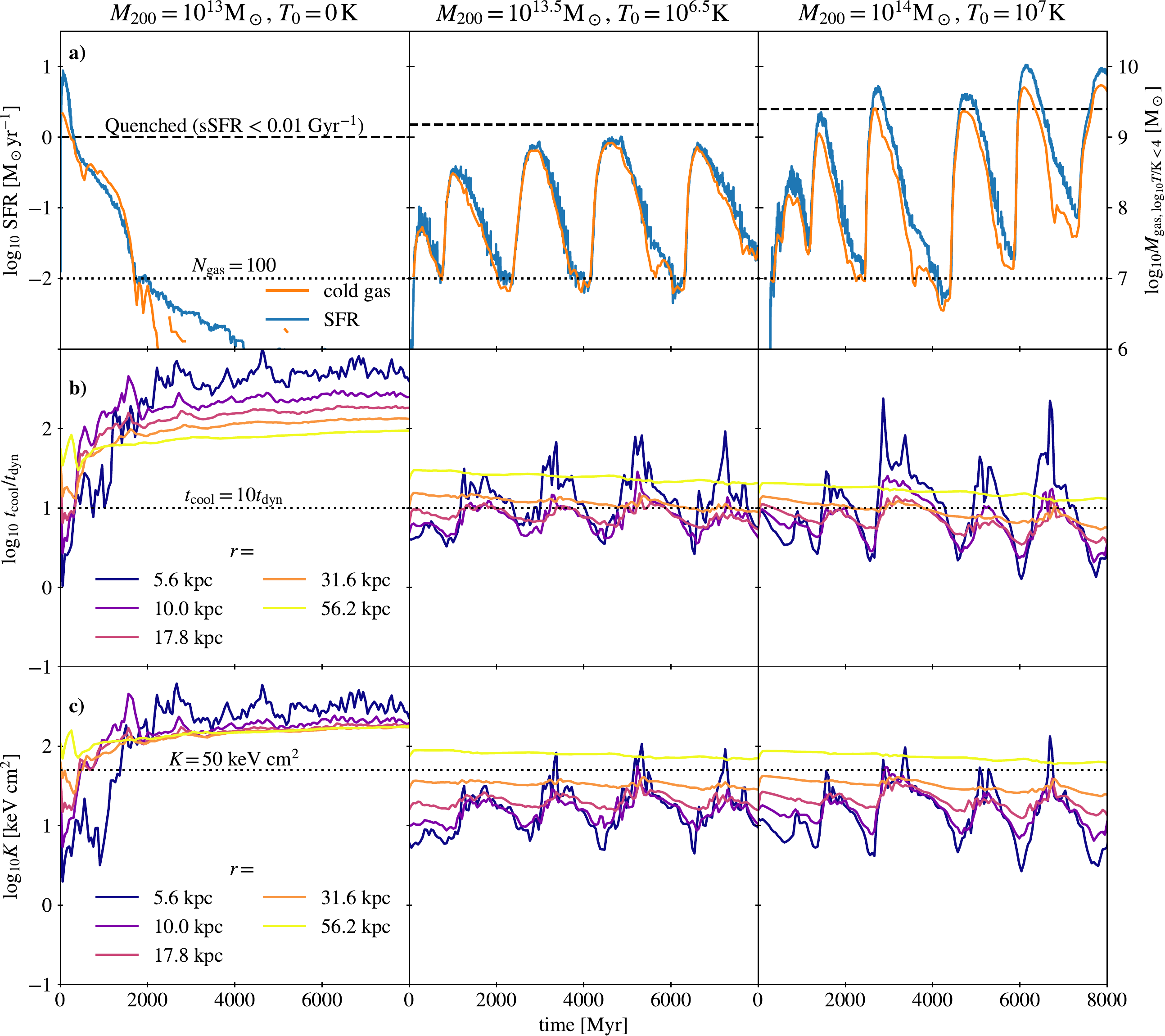}
    \caption{Comparison of the SFR and gas properties for the fiducial simulations for the three halo masses ($M_{200} = 10^{13}~\rm M_\odot$, $10^{13.5}~\rm M_\odot$, $10^{14}~\rm M_\odot$). Row (a) shows the SFR as a function of time (blue line); the dashed line corresponds to the quenching criterion of ${\rm sSFR} < 0.01~\rm Gyr^{-1}$. The orange line shows the  cold ($T_{\rm subgrid} < 10^4~\rm K$) gas mass and the mass corresponding to 100 gas particles is indicated by the dotted line. Row (b) shows $t_{\rm cool}/t_{\rm dyn}$ at different radii (differently coloured lines, as indicated in the legend), calculated within concentric shells of $\Delta \log_{10} r = 0.07$ (the dotted line indicates $t_{\rm cool}/t_{\rm dyn}=10$). Row (c) shows the evolution of the entropy (the dotted line indicates $K=50~\rm keV~\rm cm^2$, a common CC/NCC division). The low-mass halo ($M_{200} = 10^{13}~\rm M_\odot$) is quenched quickly and then maintains high entropy and large cooling times in the centre. The higher halo masses instead display oscillatory behaviour in their cooling time, $t_{\rm cool}/t_{\rm dyn}$ and entropy with the same period as the oscillations in the SFR.}
    \label{fig:thermoproperties-comparison}
\end{figure*}

\subsubsection{Central thermal feedback can regulate the ambient medium}
We now show that AGN feedback regulates star formation by raising the ratio $t_{\rm cool}/t_{\rm dyn}$, which stops the supply of cold gas. Fig.~\ref{fig:thermoproperties-comparison} shows the evolution of the SFR and cold gas (top row), $t_{\rm cool}/t_{\rm dyn}$ (middle row) and entropy (bottom row) in the ICM for our fiducial values of $T_0$for the different halo masses. We define as `cold' all gas with\footnote{Defining cold gas as $T_{\rm subgrid} < 10^{4.5}~\rm K$ gives nearly identical results.} $T_{\rm subgrid} < 10^4~\rm K$, i.e. both HI and $\rm H_2$. It is evident that the amount of cold gas directly sets the SFR. For $t_{\rm cool}/t_{\rm dyn}$ and entropy, we select gas particles within 5 narrow radial shells of width $\Delta \log r =0.07$, centred on $\{5.6, 10.0, 17.8, 31.6,56.2\}~\rm kpc$ (different colours).  The $M_{200}=10^{13}~\rm M_\odot$ halo initially has $t_{\rm cool}/t_{\rm dyn} \ll 10$ in its centre and entropies ($\la 10~\rm keV~\rm cm^2$) that are fully consistent with a CC profile. However, the initial cold gas mass in the halo is $\approx 10^{9.5}~\rm M_\odot$ and causes a high initial phase of BH accretion (Fig.~\ref{fig:SFR-BHAR-corr}). Correspondingly, AGN feedback injects a lot of energy such that after less than $2~\rm Gyr$ $t_{\rm cool}$ is raised to $\ga 10^2 \; t_{\rm dyn}$, with $t_{\rm cool} > 7~\rm Gyr$ \citep[a common CC threshold, e.g.][]{gaspari2014}, and $K \approx 10^2~\rm keV~\rm cm^2$. Hence, the initial CC profile is quickly converted into an NCC profile due to AGN feedback. The initial high BHAR produces a long $t_{\rm cool}$ that prevents any further star formation for many Gyr.

The higher-mass haloes ($10^{13.5}~\rm M_\odot$ and $10^{14}~\rm M_\odot$) on the other hand, show periodic oscillations in their $t_{\rm cool}/t_{\rm dyn}$ ratio and entropy within the inner $30~\rm kpc$, in analogy to the SFR oscillations discussed above. Their central entropy sometimes exceeds $50 ~\rm keV\,\rm cm^2$ and $t_{\rm cool}/t_{\rm dyn}$ regularly drops below $10$. Beyond $\approx 30~\rm kpc$, $t_{\rm cool}/t_{\rm dyn}$ and the entropy remain nearly constant, however. Therefore, they remain a CC cluster over most of their evolution. 

Comparing the oscillations in SFR/cold gas and in the thermodynamic ICM properties, it is apparent that the former have peaks (and troughs) slightly earlier, by a few $100~\rm Myr$. Only after the maximum SFR, when this is already declining again, are the $t_{\rm cool}/t_{\rm dyn}$ and entropy reaching their respective maxima, and analogously for minima. This indicates that the AGN regulates the ICM and induces the periodicity in the SFR. Indeed, simulations without AGN feedback (not shown) show catastrophic cooling and continuously high SFRs. 

The amount of cold gas in the most massive cluster ($M_{200}=10^{14}~\rm M_\odot$) never exceeds $10^{10}~\rm M_\odot$, and regularly drops below $10^8~\rm M_\odot$. Both trends are consistent with observations of molecular gas in BCGs \citep[e.g.][]{edge2001, salome2003, pulido2018}. A connection between cold gas and ICM state was also demonstrated observationally by \citet{pulido2018}. They only detected molecular gas ($M_{\rm mol} > 10^8~\rm M_\odot$) in CC clusters that either have central cooling time $t_{\rm cool} \lesssim 1~\rm Gyr$ or a central entropy $K \lesssim 35 ~\rm keV~\rm cm^2$. This is in good agreement with our simulation, where the cold gas mass drops below $10^8~\rm M_\odot$ only when in the centre we have $t_{\rm cool} \ga 1~\rm Gyr$ or $K \ga 35 ~\rm keV~\rm cm^2$ and vice versa.

A closer look at Fig.~\ref{fig:thermoproperties-comparison} shows that AGN feedback in the simulations raises the cooling time until the ICM is marginally stable, corresponding to $t_{\rm cool} \approx 10\, t_{\rm dyn}$, such that gas is stopped from precipitating and condensing. AGN feedback therefore mainly affects the region $r \ll 10^2$~kpc, where $t_{\rm cool} < 10\, t_{\rm dyn}$ initially. This happens quickly after the SFR and time-averaged BHAR have reached their peak values. After $t_{\rm cool}$ has been raised to $10\, t_{\rm dyn}$, the SFR and BHAR continue to decline until the remainder of the cold gas in the centre has been removed.

Suppressing the precipitation will eventually suppress the fueling of the BH. However, for the high halo masses the cooling time is smaller than $t_\text{H}$ out to $r > 10^2$~kpc (see rows d and e of Fig.~\ref{fig:BAHAMAS-comparison}). Hence, a revival of the cooling flow is unavoidable, leading to feedback cycles. The cooling time at the radius where $t_{\rm cool} \approx 10\, t_{\rm dyn}$ therefore determines the time until the next episode of cold gas condensation. This argument also explains why the cycles get shorter for higher $T_0$ (which can be seen from Fig.~\ref{fig:SFH-summary}). From Fig.~\ref{fig:BAHAMAS-comparison} (rows d and e) we see that for higher $T_0$ the radius at which the condition $t_\text{cool}=10\, t_\text{dyn}$ is reached smaller and that the cooling time at this radius is shorter. While the initial $t_{\rm cool}$ profile is increasing with $r$, after the first episode of high BHAR the $t_{\rm cool}$ profile becomes nearly constant inside $r\la r(t_\text{cool}=10\, t_\text{dyn})$, only after the injection of AGN feedback is significantly lower than its peak the halo will cool with the same $t_{\rm cool}$ for $r\la r(t_\text{cool}=10\, t_\text{dyn})$. As long as AGN feedback is significant, it will keep raising $t_{\rm cool}$ or prevents it from decreasing much in the centre.

\begin{figure}
    \centering
    \includegraphics[width=\columnwidth]{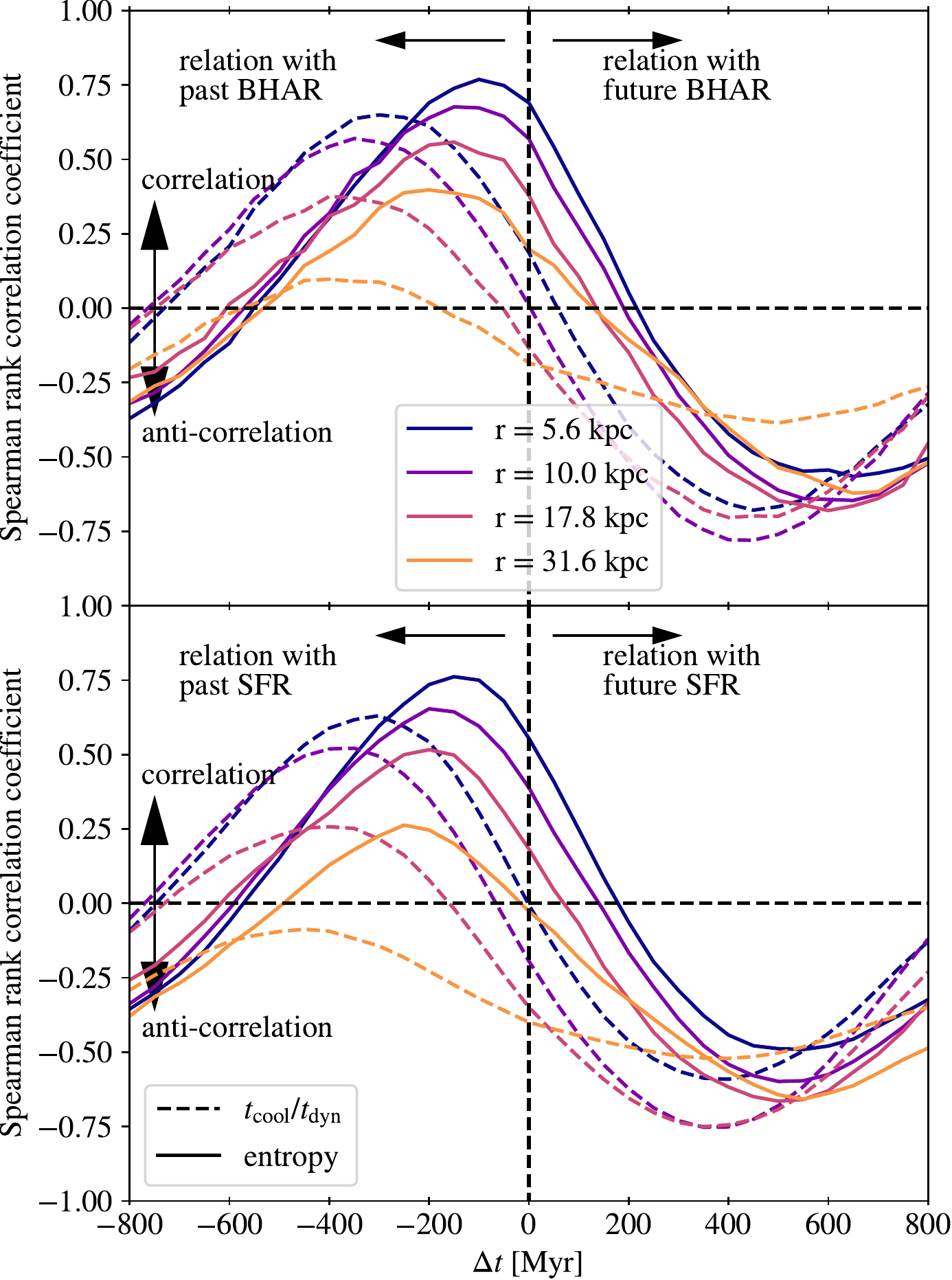}
    \caption{Top: The Spearman rank correlation coefficient for the relation between $t_{\rm cool}/t_{\rm dyn}$ (dashed curves) or ICM entropy (solid curves) and the time-averaged (over $50~\rm Myr$) BHAR (top) or instantaneous SFR (bottom) averaged as a function of the delay $\Delta t$ between the two (see text) for the $M_{200} =10^{14}~\rm M_\odot$ halo. Negative (positive) $\Delta t$ correspond to correlations of entropy and $t_{\rm cool}/t_{\rm dyn}$ with the past (future) BHAR. Different colours indicate different radii. Low $t_{\rm cool}/t_{\rm dyn}$ cause gas to precipitate into cool clouds ($t\sim 3 \times 10^3 ~\rm Myr$) that fall towards the halo centre and increase the SFR and BHAR. The increased BHAR produces outflows that raise the entropy first (within $t\sim 100-250~\rm Myr$) followed by an increase in the cooling time ($t\approx 200~\rm Myr$) that suppresses BHAR and SFR until $t_{\rm cool}$ has decreased enough for the next cycle.
    }
    \label{fig:correlation-BHAR}
\end{figure}

\subsubsection{The causal link between the ICM, AGN, and star formation}

At this point, we have shown that our massive-halo simulations ($10^{13.5}$ and $10^{14} ~\rm M_\odot$) exhibit cyclic variations in the SFR which are strongly correlated (although not perfectly synchronized) with long-term BHAR variations. Likewise, key thermodynamic ICM properties show cyclic variations with a similar period, at least in the central $\la 30 ~\rm kpc$. This naturally raises the question of how these different properties are causally connected: which variations are the drivers, and which are the result of changes in the AGN activity? To address this question, we use the same time-shifted correlation technique that we have already employed for Fig.~\ref{fig:SFR-BHAR-corr}, but now we apply it to the relation between time-averaged BHAR or SFR on the one hand, and the ICM properties $t_{\rm cool}/t_{\rm dyn}$ or entropy on the other.

For each of the four resulting pairs of variables (BHAR-$t_{\rm cool}/t_{\rm dyn}$, BHAR-entropy, SFR-$t_{\rm cool}/t_{\rm dyn}$, SFR-entropy), we plot in Fig.~\ref{fig:correlation-BHAR} the time-shifted Spearman rank correlation coefficient as a function of the shift $\Delta t$ by which we offset the BHAR and SFR evolutions. The correlations with BHAR are shown in the top panel, those with SFR in the bottom one. Since the evolutions of both $t_{\rm cool}/t_{\rm dyn}$ and entropy depend strongly on radial position within the halo (Fig.~\ref{fig:thermoproperties-comparison}), we show four variants of each correlation, with the ICM properties measured within radial shells of width $\Delta \log r = 0.07$ centred on radii $\{5.6, 10.0, 17.8, 31.6\} ~\rm kpc$ (different colours). As in Fig.~\ref{fig:SFR-BHAR-corr}, the feature of key interest is the time shift ($\Delta t$) of the peak (anti-/)correlation as well as its strength: a high - positive or negative - peak at small positive $\Delta t$ indicates that the ICM quantity under consideration will have a strong impact on the BHAR in the near future, whereas a high peak at small negative $\Delta t$ indicates that the ICM quantity has been strongly affected by the recent BHAR. Peaks with a lower amplitude correspond to a weaker connection, whereas peaks at a larger (absolute) $\Delta t$ indicate that the connection has a longer delay time.

Focusing first on the correlation of BHAR with entropy (top panel, solid lines), we see a strong positive correlation coefficient at mildly negative time shifts $\Delta t$. Both the location of the peak and its height vary systematically with radius, in the sense that the ICM closest to the BH ($r=5.6~\rm kpc$, indigo) shows the strongest positive peak correlation with the shortest time offset ($r_{\rm s} \approx 0.75$ and $\Delta t \approx -10^2 ~\rm Myr$, respectively). At larger radii, the correlation is both weaker and has a longer delay ($r_{\rm s} \approx 0.45$ and $\Delta t \approx -250 ~\rm Myr$, respectively, for $r = 31.6 ~\rm kpc$). The peak anti-correlation has a much larger $\Delta t$ ($\approx + 600 ~\rm Myr$), with no strong radial dependence. In other words, an increase in BHAR is rapidly followed by an inside-out increase of the ICM entropy --- consistent with the qualitative picture from Fig.~\ref{fig:phases-temp} --- while the driving effect of low entropy on BHAR takes much longer to develop.

The situation is broadly similar for the correlation between BHAR and $t_{\rm cool}/t_{\rm dyn}$ (top panel, dashed lines). Here, both the positive and negative peaks are offset by approximately $\pm 4\times 10^2 ~\rm Myr$; the former shows the same systematic trend as for entropy (weaker peak correlations at more negative $\Delta t$ for larger radii), with overall larger shifts and lower peak correlation strengths. This implies that AGN feedback affects ICM cooling less directly than it influences entropy, but also that gas cooling has a more significant and faster impact on future BH activity than a decreasing entropy has. 

We note that, contrary to the positive correlation peaks, the $t_{\rm cool}/t_{\rm dyn}$ negative correlation peaks show no strong systematic ordering with radius (except for the largest bin at $\approx 30 ~\rm kpc$, which shows very little cooling time variation in any case; Fig.~\ref{fig:thermoproperties-comparison}). As discussed above, gas cooling only affects the BH after some delay to account for its migration towards the centre, and this delay should be larger for gas cooling from larger radii. The fact that we do not see such a differential delay may imply that only gas cooling from relatively close to the centre (within a few kpc) has a significant impact on the BHAR.

The correlations with SFR (bottom panel of Fig.~\ref{fig:correlation-BHAR}) broadly mirror those for BHAR, with a general shift of all features by $\sim -10^2 ~\rm Myr$, as expected from Fig.~\ref{fig:SFR-BHAR-corr}. This suggests that star formation itself does not significantly influence the ICM (e.g. through SN feedback), and merely acts as an `indicator' of imminent AGN feedback. Also consistent with Fig.~\ref{fig:SFR-BHAR-corr} is the shorter positive time shift of the peak SFR-$t_{\rm cool}/t_{\rm dyn}$ anti-correlation: ICM conditions more conducive to cooling first lead to an increase in SFR (within $\approx 3\times 10^2 ~\rm Myr$), and only later to a higher BHAR.

In summary, we find the following picture of the connection between the ICM, AGN, and SFR in our simulated cluster. Decreasing cooling times lead to low $t_{\rm cool}/t_{\rm dyn}$ causing gas to precipitate into cool clouds (on a timescale of $\sim 3\times 10^2 ~\rm Myr$) that fall towards the halo centre --- mostly from the innermost ICM --- with a further $\sim 10^2~\rm Myr$ delay. There they boost the BHAR, which leads to (anisotropic) outflows of gas that raise the ambient entropy from the inside out (within $\approx 100-250 ~\rm Myr$), and subsequently --- as the bubbles disperse and cover a larger volume fraction --- increase the ICM cooling time. AGN feedback proceeds until $t_{\rm cool}> 10 \, t_{\rm dyn}$ and precipitation is suppressed. Long cooling times suppress both the BHAR and SFR, until $t_{\rm cool}/t_{\rm dyn}$ has dropped sufficiently to initiate the next cycle, which happens after a time similar to the cooling time at the radius out to which precipitation was halted. Any positive feedback effects from the AGN --- i.e. outflows triggering precipitation via a reduction in $t_{\rm cool}$ --- are subdominant and/or occur on much longer time-scales than the negative feedback due to the increase of entropy of the inner ICM.

Fig.~\ref{fig:thermoproperties-comparison} shows that for the larger halo masses (i.e. $10^{13.5}~\rm M_\odot$ and $10^{14}~\rm M_\odot$) the entropy and cooling time do not evolve much beyond $30~\rm kpc$, which is the precipitation radius $r(t_{\rm cool}\approx 10\, t_{\rm dyn})$. This implies that AGN feedback does not strongly influence gas outside the immediate neighbourhood of the BH. It is natural for the AGN to couple to this region since this is region where condensation occurs. However, AGN feedback could have overshot and affected larger radii, as is the case in the low-mass halo. The fact that this does not occur could be partly because the AGN feedback adopted in our simulations is intrinsically isotropic as opposed to jet-like. However, as demonstrated in Fig.~\ref{fig:entropy-mean}, this feedback still naturally leads to highly anisotropic bubble-like outflows that reach distances as far as $100~\rm kpc$ from the centre. One possibility is therefore that these buoyantly rising bubbles do not significantly impact the mean values of the entropy at $r\ga 30~\rm kpc$. It is possible that our AGN feedback would raise the entropy  out to $r > 30~\rm kpc$ if $t_{\rm cool}/t_{\rm dyn} \la 10$ at these radii, because in that case AGN would need to raise the cooling time at these large radii in order to suppress condensation. Since observed CC and NCC clusters differ in their entropy profiles out to much larger radii (see Fig.~\ref{fig:entropy-mean}), in our current set-up AGN feedback alone does not convert a CC into an NCC cluster. Such a transformation then requires some form of additional input: for example, a major merger may generate significant entropy through shock-heating, and/or may cause the BH to obtain enough fuel to heat the complete core. Unlike the behaviour of the high-mass haloes, AGN feedback is evidently able to convert the low-mass halo ($10^{13}~\rm M_\odot$)  from a CC into an NCC cluster.

%% file: precipitation.tex
\subsection{Comparison with the precipitation framework} \label{subsec:precip}

The causal link between the ICM, AGN and star formation agrees with the precipitation framework \citep[e.g.][]{voit2017}. The evolution of the ratio $t_{\rm cool}/t_{\rm dyn}$ determines the new episodes of star formation as expected.

\begin{figure}
    \centering
    \includegraphics[width=\columnwidth]{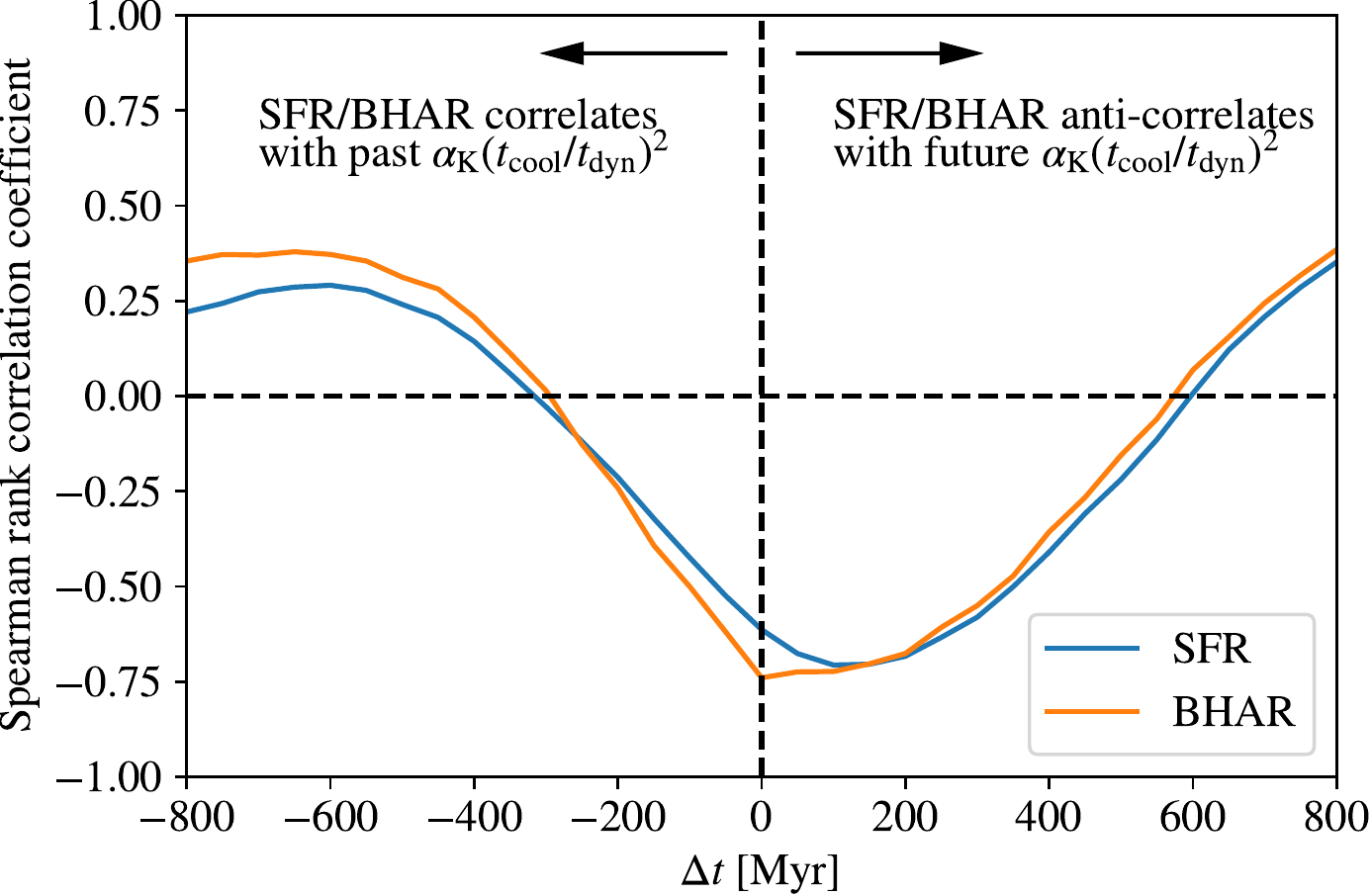}
    \caption{The Spearman rank correlation coefficient between SFR (BHAR) and volume-weighted $\alpha_{\rm K} (t_{\rm cool}/t_{\rm dyn})^2$ as a function of the delay time $\Delta t$ between the two, for the $M_{200}=10^{14}~\rm M_\odot$ halo. Negative values of $\Delta t$ show how the SFR or BHAR correlate with the past $\alpha_{\rm K} (t_{\rm cool}/t_{\rm dyn})^2$ and vice versa. There is no strong correlation between the past $\alpha_{\rm K} (t_{\rm cool}/t_{\rm dyn})^2$ and the current SFR or BHAR, in disagreement with the precipitation framework.}
    \label{fig:slope-voit-corr}
\end{figure}

\subsubsection{Entropy slope-AGN connection}
One of the predictions of the precipitation framework \citep[e.g.][]{voit2017} is that gas will precipitate when it satisfies inequality (\ref{eq:voit}) for the logarithmic entropy slope $\alpha_{\rm K}$. To check if this is the case in our simulations, we calculate $\alpha_{\rm K} (t_{\rm cool}/t_{\rm dyn})^2$ and show a time-shifted correlation with the BHAR and SFR in Fig.~\ref{fig:slope-voit-corr}. We calculate the volume-weighted entropy slope using all hot gas particles ($T>10^{5.5}~\rm K$) between $3~\rm kpc$ and $30~\rm kpc$. Based on the precipitation framework we expect that BHAR and SFR anti-correlate with past $\alpha_{\rm K} (t_{\rm cool}/t_{\rm dyn})^2$: short cooling times should lead to high BHAR and SFR. Instead, Fig.~\ref{fig:slope-voit-corr} shows the opposite: the peak anti-correlation occurs at slightly positive $\Delta t$. This indicates that the impact of AGN feedback on the entropy slope dominates the correlations. Our picture therefore is that AGN feedback causes a phase of decreased $\alpha_{\rm K} (t_{\rm cool}/t_{\rm dyn})^2$, followed by a decreasing SFR and BHAR on timescales of around $500~\rm Myr$ and  due to the high $t_{\rm cool}/t_{\rm dyn}$ a maximum in  $\alpha_{\rm K} (t_{\rm cool}/t_{\rm dyn})^2$ is reached. This is followed by a period of high SFR and BHAR while the slope is decreasing on timescales of $\Delta t \approx 500~\rm Myr$. 

\begin{figure}
    \centering
    \includegraphics[width=.97\columnwidth]{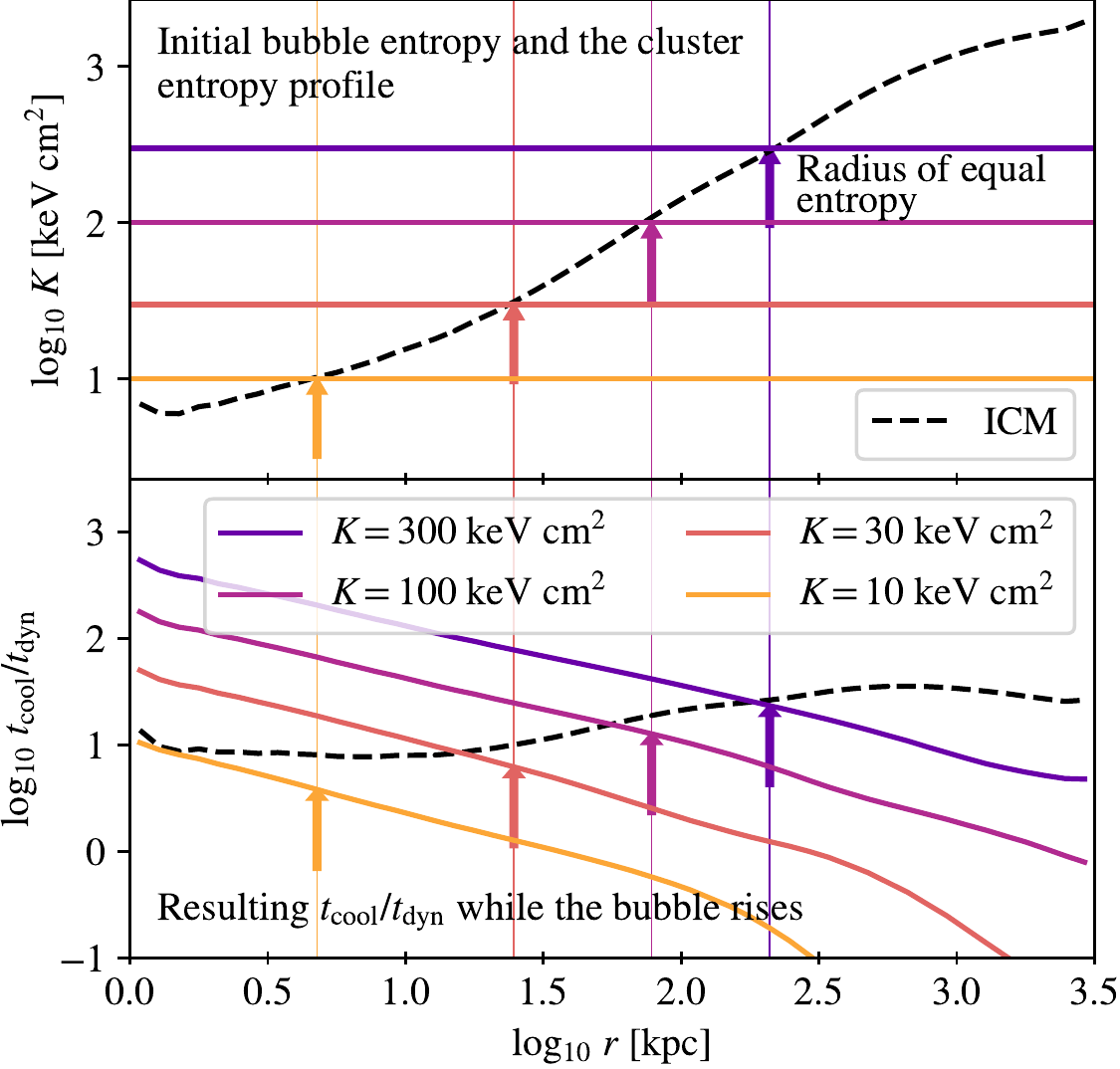}
    \caption{Adiabatic evolution of a gas bubble rising from the cluster centre ($M_{200}=10^{14}~\rm M_\odot$) with different initial entropies (different colours). In the top panel we compare the initial bubble entropies with the entropy of the ICM and indicate the radii where the two are equal by an arrow and a thin line. The bottom panel shows the cooling time over dynamical time calculated using the \citet{ploeckinger2020} cooling tables. High-entropy bubbles need to rise to radii significantly larger than the radius where their entropy matches that of their environment in order to obtain $t_{\rm cool}/t_{\rm dyn} \ll 10$, as required to condense and precipitate.
    }
    \label{fig:bubble-K}
\end{figure}

\subsubsection{Does adiabatically expanding gas become unstable?}
To determine whether gas heated by feedback will start condensing and precipitate back towards the BH, we consider the idealised situation of having a recently heated feedback bubble moving buoyantly to larger radii, until it reaches the radius at which the entropy within the bubble is the same as that of the surrounding medium. The bubble expands until it reaches the pressure of its surroundings. From there on, it will rise buoyantly while satisfying $P_{\rm bubble}=P_{\rm ICM}(r)$, where $P_{\rm bubble}$ is the internal pressure of the bubble and $P_{\rm ICM}(r)$ is the volume-weighted average pressure of the ICM at radius $r$. The entropy injected by AGN (or stellar) feedback is given by
\begin{align}
    K &= 125 ~{\rm keV} ~{\rm cm^2} \left( \frac{\Delta T}{10^{8.5}~\rm K} \right) \left( \frac{n_{\rm H}}{0.1 ~\rm cm^{-3}}\right)^{-2/3},
\end{align}
where $\Delta T$ is the temperature increase due to feedback, $n_{\rm H}$ is the hydrogen number density of the gas that is heated, and we assume $\Delta T \gg T$, the temperature of the gas before heating. Under the assumption that the bubble remains intact, in pressure equilibrium, and rises adiabatically (i.e. the entropy is constant), we can determine its density and temperature at any radius:
\begin{align}
    n_{\rm H} &= \left[ \frac{P_{\rm ICM}(r)}{K} \right]^{3/5}, \\
    T &= \frac{P_{\rm ICM}(r)^{2/5} K^{3/5}}{k_{\rm B}}.
\end{align}
Using these equations, we can compare the ratio $t_{\rm cool}/t_{\rm dyn}$ in the rising feedback bubble with that of the gas at the same radius, using the cooling times from \citet{ploeckinger2020}. Fig.~\ref{fig:bubble-K} shows the entropy and $t_{\rm cool}/t_{\rm dyn}$ for the buoyantly rising gas in the $M_{200}=10^{14}~\rm M_\odot$ halo. We consider bubble entropies in the range $10-300 ~\rm keV~\rm cm^2$ (different colours). After the bubble has risen to the radius corresponding to its own entropy (arrows and vertical thin lines), its $t_{\rm cool}/t_{\rm dyn}$ ratio is almost identical to that of the surrounding gas, which has $t_{\rm cool}/t_{\rm dyn} \ga 10$ (bottom panel). This implies that gas that rises buoyantly to the radius matching its own entropy will not have  $t_{\rm cool}/t_{\rm dyn} \ll10$ and hence will not condense. 

This toy model applies equally to AGN and SN feedback. However, bubbles produced by AGN feedback will typically attain higher entropies because the heating temperature $\Delta T$ is higher. Figs.~\ref{fig:phases} and \ref{fig:phases-temp} show that lower temperature gas ($T\approx 10^{7.2}~\rm K$) can be brought to significantly larger radii than in our toy model ($r\approx 100~\rm kpc$). If gas with entropies around $20~\rm keV~\rm cm^2$ is brought to higher altitude (where $t_{\rm dyn}$ is larger), then this gas will have $t_{\rm cool}/t_{\rm dyn}<10$. This means that cold gas lifted by hotter feedback bubbles, as well as feedback bubbles that overshoot the radius with the same entropy, can produce condensation and the corresponding precipitation.

%% file: resolution.tex
\begin{figure*}
	\includegraphics[width=.97\linewidth]{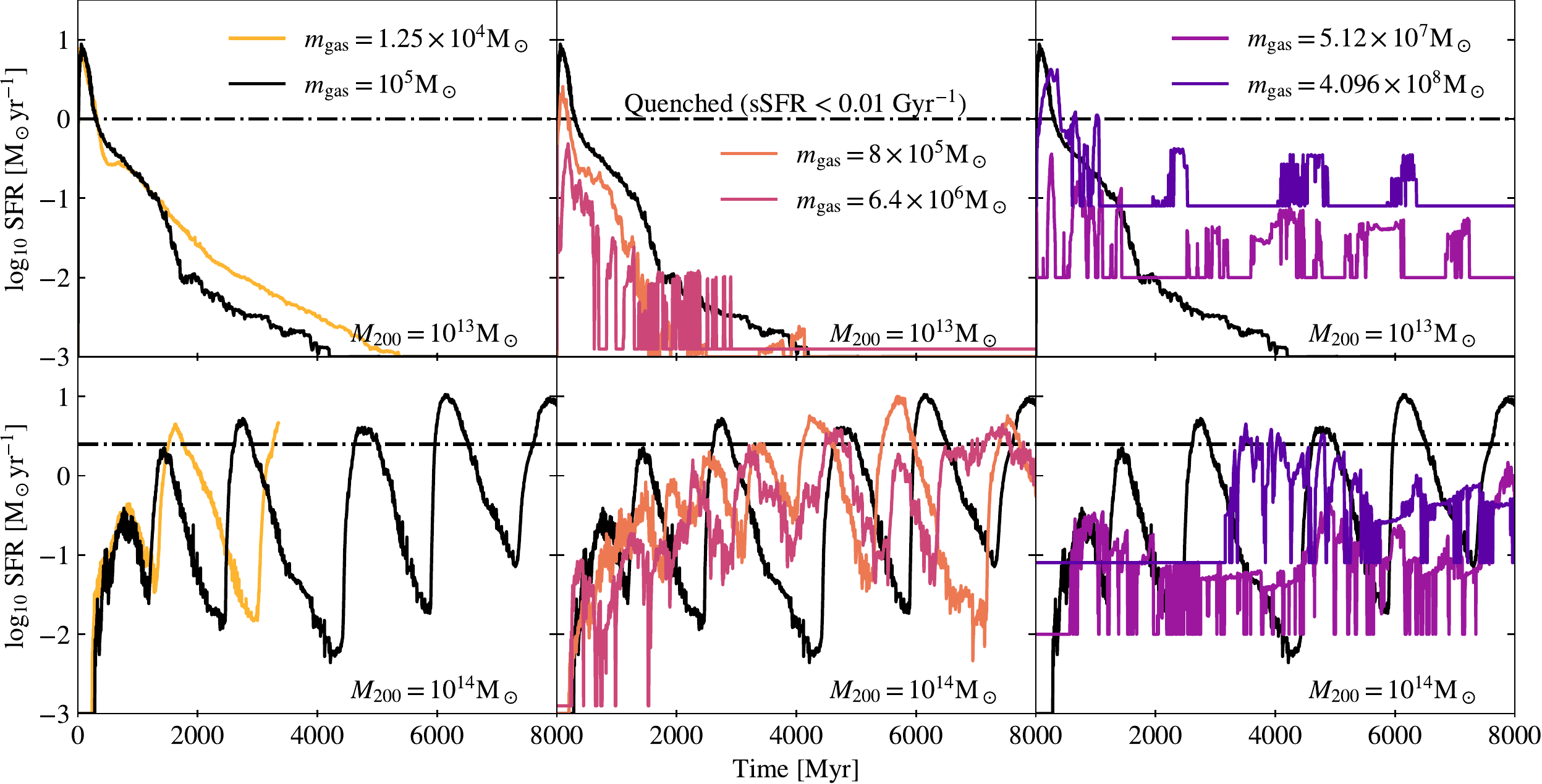}
    \caption{Comparison of the star formation history (SFH) for the $M_{200} =10^{13}~\rm M_\odot$ (top row) and $10^{14}~\rm M_\odot$ (bottom row) halo at different resolutions (different columns and colours). The horizontal black dash-dotted line shows the quenching criterion of sSFR $=10^{-2}~\rm Gyr^{-1}$ for the initial stellar mass of the BCG. The low-mass halo ($M_{200}=10^{13}~\rm M_\odot$) is well converged across all resolutions. The highest mass halo ($M_{200}=10^{14}~\rm M_\odot$) converges only up to a resolution of $6.4\times 10^6~\rm M_\odot$; lower resolutions do not show oscillatory behaviour. For the lowest resolutions, the SFR is unresolved during some times; in this case, we set the minimum SFR to $1/8 \times 10^{-2}, 10^{-2}$ and $8 \times 10^{-2} ~\rm M_\odot ~\rm yr^{-1}$ for the resolutions of $m_{\rm gas} = 6.4 \times 10^6, 5.12\times 10^7$ and $4.096\times 10^8~\rm M_\odot$, motivated by equation (\ref{eq:SFRmin}).}
    \label{fig:SFH-res}
\end{figure*}

\subsection{What resolution is required?} \label{subsec:resolution}

To determine what numerical resolution is required to capture the interaction between the ICM and the BH, we compare sets of simulations for the lowest ($M_{200} = 10^{13}~\rm M_\odot$) and highest ($M_{200} = 10^{14}~\rm M_\odot$) halo mass\footnote{The $10^{14}~\rm M_\odot$ halo at the highest resolution was not run for long due to its computational expense.}. The particle mass is varied by factors of 8 and the gravitational softening by factors of 2. The highest-resolution simulation has a particle mass of $1.25\times 10^4~\rm M_\odot$, 8 times lower than our fiducial simulation and comparable to high-resolution zoom simulations of individual galaxies \citep[e.g.][]{font2020}. The lowest-resolution simulation has a particle mass of $4\times 10^{8}~\rm M_\odot$, which is comparable to the resolution used in cosmological simulations of large samples of galaxy clusters \citep[e.g.][]{mccarthy2017,barnes2017b,cui2018}. 

The top and bottom rows of Fig.~\ref{fig:SFH-res} show the SFH of the $M_{200}=10^{13}~\rm M_\odot$ and the $M_{200}=10^{14}~\rm M_\odot$ haloes, respectively. We do not expect perfect convergence for several reasons. First, the initial density profile is realised using a finite number of particles based on stochastic sampling. Second, stellar and AGN feedback are implemented stochastically. Third, the density threshold for star formation implies the existence of a minimum nonzero SFR,

\begin{equation}
{\rm SFR}_{\rm min} = 3.5 \times 10^{-3} ~ {\rm M_\odot} ~{\rm yr}^{-1} \left( \frac{m_{\rm gas}}{10^7~\rm M_\odot} \right) \left( \frac{n_{\rm H}}{0.1 ~\rm cm^{-3}}\frac{T}{8\times 10^3~\rm K} \right)^{0.2}. \label{eq:SFRmin}
\end{equation}
Taking these considerations into account, we see good convergence for the $10^{13}~\rm M_\odot$ halo. In particular, for all resolutions the SFR declines rapidly to values significantly below the division between star-forming and quenched galaxies (horizontal line). For the $10^{14}~\rm M_\odot$ halo, the convergence is good down to a particle mass of $6.4\times 10^6~\rm M_\odot$. For higher particle masses (i.e. lower resolutions) the SFH no longer shows oscillatory behaviour. This can be understood by the fact that at resolutions of $5.12\times 10^7~\rm M_\odot$ and $\varepsilon=2.4~\rm kpc$ the central region around the BH is not properly resolved and therefore the ICM-AGN connection is no longer modelled correctly. 

Unlike the resolution, randomness in the simulation produced by our random number generator does not qualitatively impact our results (see Appendix \ref{app:random}).

%% file: comparison.tex
\subsection{Comparison with previous simulations} \label{subsec:previous-sims}

\subsubsection{Different AGN feedback models}
We have shown that our simulations do not have a persistent cooling flow. The BH limits its own growth as well as that of the galaxy by regulating the properties of the ICM. This is accumplished with AGN feedback injected in purely isotropic and thermal form. We now compare these results with previous findings in the literature that used different AGN feedback models. \citet{gaspari2011} showed that an AGN model that uses Bondi-Hoyle accretion and mechanical AGN feedback is able to prevent the formation of a cooling flow in haloes of mass $\approx 4 \times 10^{13}~\rm M_\odot$. They find that this is also true for a model in which AGN feedback is triggered by the accretion of cold gas within $3~\rm kpc$ from the BH. \citet{gaspari2013} find that their cold gas triggered AGN feedback model produces chaotic accretion of gas onto the BH. \citet{gaspari2014} find that self-regulated mechanical jets produce both a realistic cooling time and $L_{\rm X} - T_{500}$ relation, while thermal blast models that are not self-regulated cause excessively long cooling times in the centres of galaxy cluster and too low $L_X$ for $T_{500} \la 1~\rm keV$. Our study agrees with these results in that self-regulated simulations are able to produce realistic cooling times in the centres of galaxy clusters, but also highlights that this can be achieved with purely thermal feedback. 

\citet{meece2017} show that in general the precise implementation of AGN feedback is unimportant and that AGN will self-regulate except for the case of purely thermal AGN feedback. At first sight this appears to contradict our findings. Like us, \citet{meece2017} use a variant of the \citet{booth2009} implementation of thermal AGN feedback. However, \citet{meece2017} use a heating temperature of only $10^7~\rm K$, which is significantly below the virial temperature of their halo and which does not satisfy the \citet{dalla-vecchia2012} criterion for preventing numerical overcooling. We use a significantly higher heating temperature of $10^{8.5}~\rm K$ to suppress such spurious energy losses. As long as both conditions are satisfied the choice of $\Delta T$ will have a limited impact on the cyclic behaviour of the haloes. However, the exact $t_{\rm cool}$ in the centre and mass loading of the outflows will differ. Altogether this sketches a scenario that if the BH model allows self-regulation and if the feedback implementation does not suffer from excessive numerical overcooling, then the exact implementation of the AGN feedback is not critical.

\subsubsection{High-mass clusters like Perseus}
In the literature, idealised clusters like Perseus with halo masses of around $6\times 10^{14}~\rm M_\odot$, have been extensively studied \citep{li2014a,li2014b,li2015,li2017}, so far mostly using simulations with adaptive mesh refinement (AMR). E.g. \citet[][]{li2015} show that the cold gas mass is regulated by AGN feedback and that it sets both the SFR and average BHAR, with a cycle between episodes of high and low SFRs. They also show (their fig. 2) that on time scales below $t<10^2~\rm Myr$ the BHAR varies over at least 3 orders of magnitude. These findings are in excellent agreement with ours. However, they predict significantly higher SFRs, implying that on average the BCG is actively forming stars and is only quenched for short periods of time. Following cycles of high SFR, the cold gas mass decreases slower than what we are finding. This is probably linked to the fact that a ring of cold gas is formed around the BH that is difficult to destroy by their jet feedback. The mass of the cold ring therefore decreases almost with the speed at which it is converted into stars. We also obtain a cyclic behaviour with episodes of low and high SFRs/BHARs, but in our case the BCG remains quenched during nearly the complete evolution. Similar to \citet{li2015}, our simulations form a ring of cold gas, but in our case it is quickly disturbed. Note however, that our most massive galaxy cluster is almost $1~\rm dex$ lower in halo mass than that in \citet{li2015}.

Cyclic behaviour is also predicted by the simulations of \citet{prasad2015} who report an increase in the BHAR when $t_{\rm cool}/t_{\rm dyn} \leq 10$ due to the condensation of cold gas that feeds the BH. Similar to our findings, their AGN feedback is not able to convert the CC cluster into an NCC cluster. They also find a rotationally supported cold ring around the centre of the BH. The reason that we do not find such a long-lived cold gas ring is likely because our AGN feedback injection is isotropic. \citet{qiu2019} also find that there is a regulation between the cold gas and the BHAR. They find that $t_{\rm cool}/t_{\rm dyn}$ regulates to values slightly higher than $10$ in both a Perseus-like cluster and lower-mass clusters, which is in agreement with our simulations. 

In addition to idealised cluster simulations there are simulations like C-EAGLE \citep{barnes2017,bahe2017} that model galaxy clusters in a cosmological context using nearly the same model for AGN feedback as used here. \citet{barnes2017} find that clusters in C-EAGLE have too large entropy cores and lack CC systems. \citet{barnes2018} studied the evolution of CC clusters in IllustrisTNG. They find that their CC fraction evolves from mainly CC at low $z$ to mainly NCC at high $z$, and that the evolution of the CC and NCC distribution is too strong in the simulation. 

\subsubsection{Lower-mass clusters}
Besides Perseus-mass clusters, \citet{qiu2021} simulated a group with a mass of $\approx 3 \times 10^{13}~\rm M_\odot$. They predict values of $t_{\rm cool}/t_{\rm dyn}$ similar to ours, such that the BH self-regulates. Contrary to \citet{qiu2021}, we find that clusters with this mass also show periodic behaviour in cold gas. The simulations from \citet{wang2019}, who modelled the  elliptical galaxies of NGC5044 and NGC 4472 show that also at halo masses of $M_{200} = 4\times 10^{13}~\rm M_\odot$ there is a periodic behaviour in the amount of cold gas in the centre, regulated by the ratio $t_{\rm cool}/t_{\rm dyn}$. Another different idealised cluster simulation has been performed by \citet{su2021}, who use the FIRE-2 model \citep{hopkins2018} and a dark matter halo mass of $8.5\times 10^{13}~\rm M_\odot$. They find that whether galaxies quench or not depends on the amount of energy that is injected through their different AGN feedback channels, which may however be due to the fact that their AGN model is not self-regulating. \citet{prasad2020b} studied the analytic AGN feedback `valve model' for galaxy models from \citet{voit2020} and found that their galaxies are in reasonable agreement with this model. Like us, they find that cyclic variation of the cold gas content are regulated by AGN feedback.

%% file: conclusion.tex
\section{Conclusions}
\label{sec:conclusion}
We have analysed the connection between the ICM, star formation and the AGN using hydrodynamical simulations of galaxy groups/clusters. We developed new initial conditions for a spherical equilibrium system comprising a BCG with a \citet{hernquist1990} profile, a dark matter halo with an \citetalias{navarro1997} density profile, and an ICM profile with two free parameters. The free parameter $T_0$, which sets the central temperature plateau, can be used to initialize different hydrostatic equilibrium profiles for the ICM, ranging from cool-core to non-cool-core systems. The second free parameter is used to specify the cluster gas fraction. To obtain realistic thermodynamic profiles, for each halo mass we calibrate the fiducial values for the two ICM parameters against the cosmological BAHAMAS simulation \citep{mccarthy2017}. 
Using a subgrid model similar to that used in the EAGLE simulation \citep{schaye2015}, which includes radiative cooling, star formation, BH accretion, and both stellar and AGN feedback, we simulate the evolution of haloes with $M_{200}=10^{13}~\rm M_\odot, 10^{13.5}~\rm M_\odot$ $10^{14}~\rm M_\odot$. Our main conclusions are as follows:
\begin{itemize}
    \item Low-mass galaxy clusters ($M_{200}=10^{13}\,\rm M_\odot$) behave qualitatively differently from more massive haloes. They are quickly ($t\la 1~\rm Gyr$) converted from cool-core (CC) into non-cool-core (NCC) by AGN feedback. In contrast, their more massive analogues that are initially CC show periodic behaviour in SFR, BHAR, and ICM properties (see Figs.~\ref{fig:entropy-mean},~\ref{fig:SFH-summary} and \ref{fig:thermoproperties-comparison}). The results for low-mass clusters are insensitive to their initial thermodynamic profile (as long as $t_{\rm cool}/t_{\rm dyn}$ remains realistic). A large amount of cool gas condenses towards the centre resulting in a high BHAR and AGN feedback that quickly quenches the galaxy. The evolution of higher-mass clusters is sensitive to their initial thermodynamic profile. A smaller initial $t_{\rm cool}/t_{\rm dyn}$ yields a higher SFRs that persists for most of the simulation, even after the BH has injected energy exceeding the amount needed to compensate for the initial difference.
    
    \item Even though AGN feedback is injected isotropically and thermally, its interaction with the ISM and ICM results in outflows with a biconical structure (Fig.~\ref{fig:phases-temp}). 
    
    \item The instantaneous BHAR is highly variable and has no strong correlation with the SFR. However, when averaged over timescales $t> 1~\rm Myr$, it correlates strongly with the SFR (Figs.~\ref{fig:SFR-BHAR} and \ref{fig:SFR-BHAR-corr}). 
    
    \item Before an episode of high SFR, the ratio between cooling and dynamical time ($t_{\rm cool}/t_{\rm dyn}$) drops below 10 in the centre (Fig.~\ref{fig:thermoproperties-comparison}). Overall, episodes of low entropy and small $t_{\rm cool}/t_{\rm dyn}$ are preceded by episodes of high SFR and (time-averaged) BHAR (Fig.~\ref{fig:thermoproperties-comparison}), and at the same time episodes of high/low SFR and BHAR precede episodes of high/low entropy and $t_{\rm cool}/t_{\rm dyn}$ in the centre. 
    
    \item In higher-mass clusters showing cyclic behaviour, AGN feedback influences mostly the gas within the radius where $t_{\rm cool}/t_{\rm dyn} < 10$ before the feedback episode. In this region the AGN raises the entropy until $t_{\rm cool}/t_{\rm dyn} \approx 10$, thus suppressing precipitation (Fig.~\ref{fig:thermoproperties-comparison}). At that point cold gas is however still present in the centre, which is either removed by AGN feedback or converted into stars. The cooling time just outside the region where the ratio $t_{\rm cool}/t_{\rm dyn}$ was raised to $\approx 10$ (see row d of Fig.~\ref{fig:BAHAMAS-comparison}) sets the time until the following episode of high SFR and BHAR will take place.
    
    \item Using time-shifted correlations, we quantify the delay between on the one hand SFR or BHAR, and on the other hand entropy or $t_{\rm cool}/t_{\rm dyn}$ (Fig.~\ref{fig:correlation-BHAR}). The results are in good agreement with the precipitation framework in which the value of $t_{\rm cool}/t_{\rm dyn}$ and the entropy of the ICM regulate the inflow of gas towards the centre. However, the AGN feedback influences the logarithmic entropy slope therefore it is not a good predictor for precipitation (Fig.~\ref{fig:slope-voit-corr}).  
    
    \item In order for high-entropy ($K \ga 10^2 ~\rm keV~\rm cm^2$) bubbles to obtain $t_{\rm cool}/t_{\rm dyn}<10$ and condense, they are required to reach significantly larger radii than the radius at which their entropy matches that of the surrounding ICM (Fig. \ref{fig:bubble-K}). Our simulations show that high-entropy bubbles bring cold gas to larger radii such that the cold gas can condense, probably due to pushing of cold gas by hot bubbles, afterwinds caused by hot bubbles and/or bubbles overshooting their equilibrium radius.
    
    \item We find good convergence with the numerical resolution over more than four orders of magnitude in gas particle mass (Fig.~\ref{fig:SFH-res}). However, the oscillatory behaviour of the high-mass clusters is no longer reproduced for gas particle masses below $\approx 6 \times 10^6~\rm M_\odot$, when the interaction between the central ICM and the AGN ceases to be adequately resolved.
\end{itemize}

We have demonstrated that a subgrid model for AGN feedback that injects energy in purely thermal form can convert low-mass clusters from CC to NCC systems, while producing cyclic behaviour for higher-mass CC clusters. In the latter case, long periods during which star formation is quenched are followed by short periods of elevated SFRs and BHARs. The precipitation framework can explain the connection between the ICM and the central galaxy found in our simulations. In the future we intend to investigate how sensitive these results are to the models for BH accretion and AGN feedback.   

%% file: seed.tex
\section{The impact of random numbers}
\label{app:random}

\begin{figure}
	\includegraphics[width=\columnwidth]{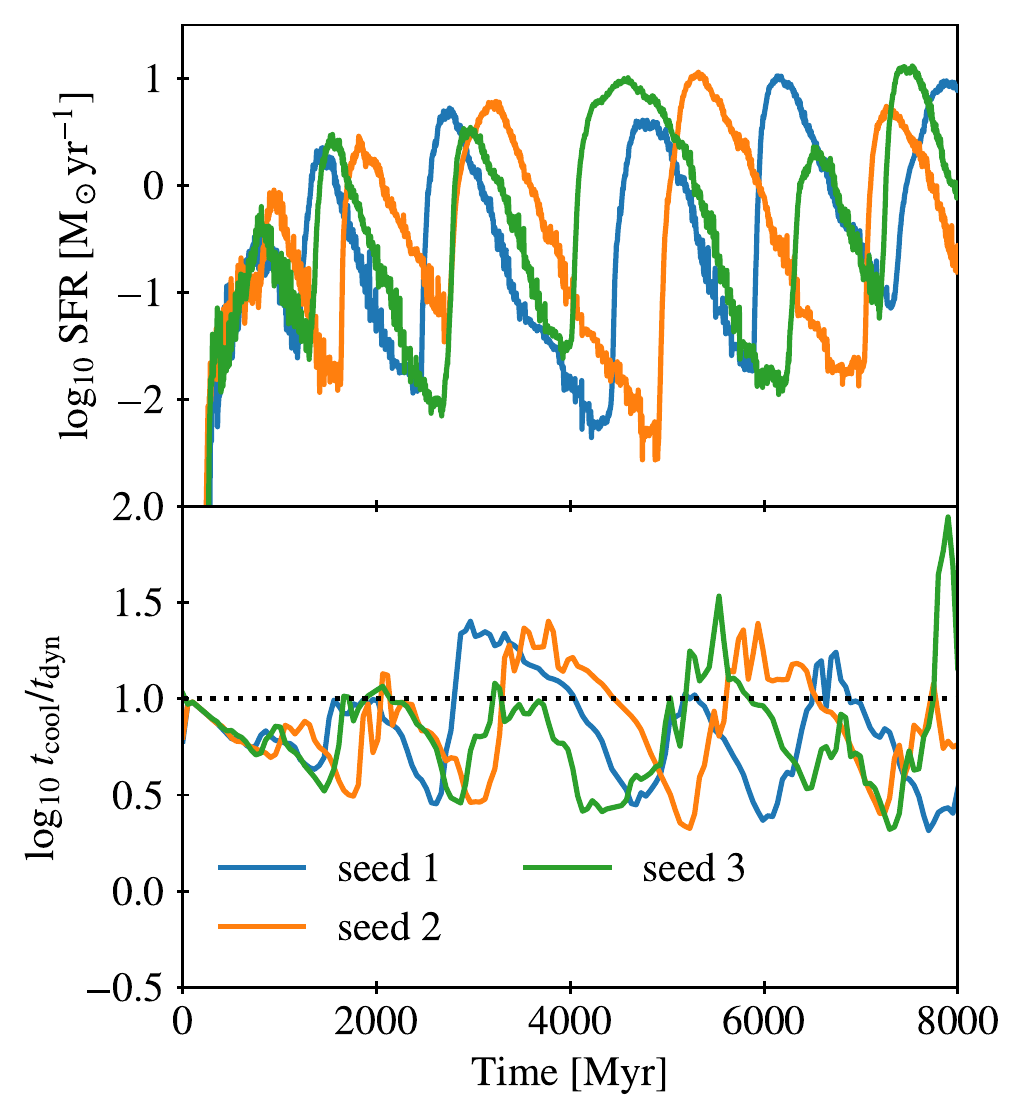}
    \caption{Comparison of the SFR (top panel) and $t_{\rm cool}/t_{\rm dyn}$ ratio (bottom panel) for the $10^{14}~\rm M_\odot$ halo using three different random seeds (different colours). Although the absolute periods and phases are different for different seeds, the relation between the SFR and $t_{\rm cool}/t_{\rm dyn}$ ratio remains largely unaffected.}
    \label{fig:SFH-seed}
\end{figure}

The subgrid model used in the simulations depends heavily on random numbers because the star formation and SNe/AGN feedback are stochastic. Therefore, to investigate whether our results are qualitatively influenced by different random numbers, we investigate two additional runs that have identical physics and initial conditions but use different random number seeds. Fig. \ref{fig:SFH-seed} shows the SFR and $t_{\rm cool}/t_{\rm dyn}$ ratio at 10 kpc. The exact positions and amplitudes of the peaks is influenced by the random number seeds, but the connection between the SFR and $t_{\rm cool}/t_{\rm dyn}$ ratio remains qualitatively the same. This means that the general results are insensitive to the use of different random numbers.